THE EUROPEAN
PHYSICAL JOURNAL C

# Scrutinizing various phenomenological interactions in the context of holographic Ricci dark energy models

Ehsan Sadri[1,a] , Martiros Khurshudyan[2,3,4,5,6,b], Ding-fang Zeng[7,c]

[1] Department of Physics, Central Tehran Branch, Islamic Azad University, Tehran, Iran
[2] CAS Key Laboratory for Research in Galaxies and Cosmology, Department of Astronomy, University of Science and Technology of China, Hefei 230026, People's Republic of China
[3] School of Astronomy and Space Science University of Science and Technology of China, Hefei 230026, People's Republic of China
[4] Institut de Ciencies de lEspai (CSIC), Campus UAB, Carrer de Can Magrans, s/n 08193 Cerdanyola del Valles, Barcelona, Spain
[5] International Laboratory for Theoretical Cosmology, Tomsk State University of Control Systems and Radioelectronics, 634050 Tomsk, Russia
[6] Research Division, Tomsk State Pedagogical University, 634061 Tomsk, Russia
[7] Institute of Theoretical Physics, Beijing University of Technology, Bejing 100124, China

**Abstract** In this paper, we examine two types of interacting holographic dark energy model using Pantheon supernova data, BAO BOSS DR12, CMB Planck 2015, fgas (gas mass fraction) and SZ/Xray (Sunyaev–Zeldovich effect and X-ray emission) data from galaxy clusters (GC). In particular, we considered the Holographic Ricci dark energy and Extended holographic Ricci dark energy models. During this analysis, we considered seven types of phenomenological interaction terms (three linear and four non-linear) $Q_1 = 3Hb(\rho_D + \rho_m)$, $Q_2 = 3Hb\rho_D$, $Q_3 = 3Hb\rho_m$, $Q_4 = 3Hb\left(\rho_D + \frac{\rho_D^2}{\rho_D + \rho_m}\right)$, $Q_5 = 3Hb\left(\rho_m + \frac{\rho_m^2}{\rho_D + \rho_m}\right)$, $Q_6 = 3Hb\left(\rho_D + \rho_m + \frac{\rho_D^2}{\rho_D + \rho_m}\right)$, $Q_7 = 3Hb\left(\rho_D + \rho_m + \frac{\rho_m^2}{\rho_D + \rho_m}\right)$ respectively. To find the best model we apply Bayesian Inference (BI) and use the $\Lambda$CDM as the referring model for comparison. Using the Bayesian Evidence model selection method we note that the $Q_3$ and $Q_5$ interaction terms are favored by observational data among the other ones. The obtained results also demonstrated that the evidence from the Bayesian inference method against the considered types of holographic Ricci dark energy model is strong since the $\Lambda$CDM is considered as the reference model and also the $\Lambda$CDM is preferred over the models. We also observed that the values of the deceleration parameter and the transition redshift for all models are compatible with the latest observational data and Planck 2015. In addition, we studied the jerk parameter for all models. Using our modified CAMB code, we observed that the interacting models suppress the CMB spectrum at low multipoles and enhances the acoustic peaks.

## 1 Introduction

Raised in 1998 [1], dark energy has become one of the substantial cases in modern cosmology and many models have been proposed as candidates to investigate it through the timeline of the Universe. Despite these proposed models, the dark energy is still a riddle in cosmology [2–10] (to mention a few). The cosmological constant $\Lambda$ because of its proper explanation of the Universe's expansion is the good candidate for the study of the dark energy [11–14]. In spite of this coordination, the cosmological constant suffers from some drawbacks. Lack of ability to clarify, why densities of dark energy and dark matter are of the same order while they evolve distinctly is of these drawbacks [9,15–20]. Hence the holographic dark energy (HDE) as an alternative has been proposed and drawn many attentions in recent years [21–28]. This model is originated from the holographic principle to which all of the information in a particular region of space can be drawn out from its boundary area and considered by an IR cutoff [29,30]. The energy density of HDE can be expressed as $\rho_D = 3c^2 M_p^2 / L^2$ [31–33]. In this equation, $c^2$ is a numerical constant, $M_p$ denotes the reduced Planck mass and $L$ can be taken as the size of the current Universe such as the Hubble scale [34,35].

In addition, the HDE has some problems. The holographic dark energy with event horizon leads to the causality viola-tion and choosing other cutoffs such as Hubble Horizon and particle horizon could not satisfy the accelerated expansion of the Universe [36,37]. Inspired by these problems from HDE, a model has been proposed which its length scale is the average radius of Ricci scalar curvature $|R|^{-1/2}$. This leads the dark energy density to be proportional to $R$. This is so-called the Holographic Ricci Dark Energy model (HRDE) [38]. Furthermore, the HRDE models have been extended to another model known as the Extended HRDE [39]. The HRDE can remove the fine-tuning problem and also this model avoids the causality violation and the coincidence problem [30,31,40–43].

[a] e-mail: ehsan@sadri.id.ir (corresponding author)

[b] e-mail: khurshudyan@yandex.com; khurshudyan@ustc.edu.cn; khurshudyan@tusur.ru

[c] e-mail: dfzeng@bjut.edu.cn

An approach toward avoiding the coincidence problem also is the usage of interaction term as a non-gravitational component between dark sectors [44–49] (to mention a few). In some quantum theory of gravitation, such as loop quantum gravity and Horava–Lifshitz gravity, Lorentz symmetry and general covariance are emergent features, they are usually broken at fundamental scales. Although the accelerating expansion of the universe and dark energy are phenomena on large scales, we have no definite evidence saying that the effects of such features cannot be re-organized and manifest cosmologically. Therefore, from the aspects of excluding possibilities using observations, we need to consider the interaction form seriously. In addition, it is well known that any modified/extended theory of gravity can be presented in the form of non-minimally coupled scalar field theory – the scalar field is non-minimally coupled with geometry. In this case, the interaction term will appear spontaneously in the equations describing the fluid dynamics. It is obvious that depending on the form of the non-minimal coupling we will recover different forms of interaction. On the other hand, the interaction can be introduced in a covariant way, using the velocity [50]. Simultaneously, the consideration of the interaction between the dark energy and the dark matter is a way to cross the phantom line [21,51,52]. In addition, because of the degeneracy between dark sectors in the Einstein's gravity, it could be assumed that there is a non-gravitational coupling/interaction between them which can be non-linear [53–59]. These kinds of interaction terms can be used for probing the dark energy-related problems. Hence, the cosmologists have different options for selection and comparison of linear and non-linear models.

On the other hand, the coupling between dark sectors affects the history of the Universe expansion and the framework of the structure formation [60–62]. In fact, because of the coupling, the growth of perturbations for dark matter can be highlighted and used to study the age of some celestial objects [63]. It has been also argued that this coupling between dark sectors influences the dynamical balance of crumbled framework through a procedure that the detection in the galaxy cluster Abell A586 is possible [64]. Using different observational data the constraint on the coupling between dark energy and dark matter can be more restricted which is a small and positive value. This small value of coupling constant conveys that the dark energy can decay into dark matter [65]. In addition, the cosmological observables can be affected by the existence of an interaction between dark sectors [66–70].

Moreover, the proper choice of interaction between dark energy and dark matter may have an effect on the low-$\ell$ region of the cosmic microwave background (CMB) spectrum and responsible for the CMB power spectrum suppression [63,71,72]. The nature of the dark energy and dark matter is unknown, therefore it would not be a feasible way to obtain an accurate form of the interaction from the fundamental theory and principles of physics. This might be a chief issue concerning the exploration of dark energy physics. Thus, it is wise to determine it from the phenomenological considerations [64,73–77] (to mention a few).

In this case, the phenomenological interactions have been studied in some works with holographic dark energy models. To be particular, Fu and et. al used three types of interactions ( $Q = 3Hb\rho_D$, $Q = 3Hb\rho_c$, $Q = 3Hb\,(\rho_D + \rho_c))$ in the context of holographic Ricci dark energy model (HRDE) [78]. Using SNIa, BAO, and CMB as the latest observational data they found that HRDE models are not favored by these observational data and the BIC evidence is strongly against the model. These phenomenological interactions also have been used by Li and et. al along with $Q = 3Hb\sqrt{\rho_D\rho_c}$ and $Q = 3Hb\left(\rho_m + \frac{\rho_D\rho_c}{\rho_D+\rho_c}\right)$ [79]. Unanimously, they found that $Q = 3Hb\rho_D$ is better than the other interactions in their studies. The mentioned interactions also have been studied in [80] and using SNIa, BAO, CMB and $H_0$ as the latest observational data. It was observed that $Q = 3Hb\rho_D$ and $Q = 3Hb(\frac{\rho_D\rho_c}{\rho_D+\rho_c})$ are the best models according to the results of AIC and BIC evidences.

Recently, some developments considering new forms of non-gravitational and non-linear interaction have been proposed [81]. Motivated by the aforementioned discussion, we would like to use several of these interactions and also check if these new non-linear interactions are better than the linear ones in this regards.

According to the discussion above, in this paper, we compare two models of holographic dark energy with Ricci scalar curvature namely, Holographic Ricci Dark Energy model (HRDE) and Extended Holographic Ricci Dark Energy model (EHRDE) along with seven types of interaction ($Q_1 = 3Hb\,(\rho_D + \rho_m)$, $Q_2 = 3Hb\rho_D$, $Q_3 = 3Hb\rho_m$, $Q_4 = 3Hb\left(\rho_D + \frac{\rho_D^2}{\rho_D+\rho_m}\right)$, $Q_5 = 3Hb\left(\rho_m + \frac{\rho_m^2}{\rho_D+\rho_m}\right)$, $Q_6 = 3Hb\left(\rho_D + \rho_m + \frac{\rho_D^2}{\rho_D+\rho_m}\right)$, $Q_7 = 3Hb\left(\rho_D + \rho_m + \frac{\rho_D^2}{\rho_D+\rho_m}\right)$) listed in Table 1. Using the Bayesian Evidence as a model selection tool we choose the most appropriate models among the other ones.

For investigating the cosmographical aspects of the models we use the exact function of the Hubble parameter, the deceleration parameter, and the jerk parameter [82–84]

$$H(t) = \frac{1}{a}\frac{da}{dt} = \frac{\dot{a}(t)}{a(t)}, \tag{1}$$

$$q(t) = -\frac{1}{aH^2}\frac{d^2a}{dt^2} = -1 - \frac{\dot{H}}{H^2}, \tag{2}$$

$$j(t) = \frac{1}{aH^3}\frac{d^3a}{dt^3} = q + 2q^2 - \frac{\dot{q}}{H}. \tag{3}$$

Extending these derivatives to the higher orders, for instance, one can obtain snap parameter ($s$) for $4th$ order derivative to check how the evolution of the Universe deviates from the $\Lambda$CDM [85]. In the present work, we restrict to the jerk parameter. By the use of the deceleration parameter, it is possible to check the behavior of expansion of the Universe and also its transition from the decelerated ($q > 0$) to the accelerated era ($q < 0$). In addition, we may mark the transition redshift $z_t$ (when $q = 0$), the turning point redshift between two accelerated and decelerated era. Using the cosmic jerk parameter $j$ as a dimensionless parameter we can compare the studied models with $\Lambda$CDM with $j_0 = 1$. Furthermore, a Universe with an accelerating rate of expansion has a positive value of jerk parameter.

The results of this paper for the models discussed above are based on the constraints from latest various observational datasets, namely the Pantheon Supernova type Ia, BAO from BOSS DR12, CMB from Planck 2015, and two categories of data originated from X-ray emitted from the galaxy clusters which are fgas (gas mass fraction) and SZ/X-Ray (Sunyaev–Zeldovich effect and X-ray emission) data. According to data analysis using these categories of data, the HRDE and EHRDE models remain disfavor by observational data. Also, we will see that the interaction $Q_3 = 3Hb\rho_m$ will be the best model among the other ones along with $Q_5 = 3Hb\left(\rho_m + \frac{\rho_m^2}{\rho_D + \rho_m}\right)$ as the best model in nonlinear region.

The structure of this paper is as follows. In the next section (Sect. 2) we briefly review the background equations of the models and introduce the interaction terms used in the current work. In Sect. 3, we derive the differential equations of HRDE and EHRDE models and obtain the cosmological parameters of each model according to the chosen interaction terms. In Sect. 4, the cosmographical behavior of the models has been studied. In Sect. 5, we provide the obtained results and discuss the aspects of the models. In Sect. 6, we study the behavior of the present models in the CMB angular power spectrum. The last section is allocated to some concluding remarks.

## 2 Background evolution

It is well-known that in a spatially flat FRW Universe, the Friedmann equation reads

$$3M_P^2 H^2 = \rho_D + \rho_m, \tag{4}$$

where $3M_P^2 H^2$ is the critical density and $\rho_D$ and $\rho_m$ are the density of dark energy and dark matter respectively. We may also write the dark energy and dark matter density with respect to the critical density as

$$\Omega_D = \frac{\rho_D}{3M_P^2 H^2}, \quad \Omega_m = \frac{\rho_m}{3M_P^2 H^2}, \tag{5}$$

and they obey the following relation

$$\Omega_D + \Omega_m = 1. \tag{6}$$

The consideration of interaction between dark sectors makes the energy densities of the dark energy and the dark matter to be unable to satisfy the conservation laws. Hence, this leads to the following continuity equations

$$\dot{\rho}_m + 3H\rho_m = Q, \tag{7}$$

$$\dot{\rho}_D + 3H(\rho_D + P_D) = -Q, \tag{8}$$

in which $Q$ conveys the interaction term indicating energy flow between the components. Let us consider an explicit, non-gravitational form of interaction which phenomenologically originates from the energy transfer between the dark energy and the dark matter as [48]

$$Q = 3Hbq^n\left(\rho + \frac{\rho_i\rho_j}{\rho}\right). \tag{9}$$

Where $n$ is a positive constant, $q$ is the deceleration parameter with $-1 - \frac{\dot{H}}{H^2}$ defined in Eq. (2), H is the Hubble parameter and $\rho$ would be the summation of the dark energy density and dark matter density ($\rho_D + \rho_m$). The study of Ref. [48] (which used different interacting Chaplygin gas models) shows that by choosing the sing changeable interaction originated from Eq. (9), the stable critical points and the late time attractors cannot be found, while fixing the sign of interaction the new late time attractors appears and describes, for instance, a Chaplygin gas dominated Universe. The most important achievement of Ref. [48] is the new forms of scaling attractors demonstrating new solutions of the cosmological coincidence problem. For more details, we refer the readers to Ref. [48]. In our study, we choose the fixed sign of the interactions in the Eq. (9) during the whole evolution of the Universe by consideration of $n = 0$. These kinds of interaction are very common types of non-linear interaction (as we see, for instance, in Refs. [79,86–88]). In this paper, we will consider four non-linear terms of the interaction along with 3 linear ones which are listed in the Table 1.

**Table 1** List of linear interactions ($Q_{1,3}$) and non-linear interactions ($Q_{4,7}$) considered in this paper

| Mark | Interaction | References |
|------|-------------|-----------|
| $Q_1$ | $3Hb\left(\rho_D + \rho_m\right)$ | [21,44,89–92] |
| $Q_2$ | $3Hb\rho_D$ | [21,44,89–92] |
| $Q_3$ | $3Hb\rho_m$ | [21,44,89–92] |
| $Q_4$ | $3Hb\left(\rho_D + \frac{\rho_D^2}{\rho_D+\rho_m}\right)$ | [48] |
| $Q_5$ | $3Hb\left(\rho_m + \frac{\rho_m^2}{\rho_D+\rho_m}\right)$ | [48] |
| $Q_6$ | $3Hb\left(\rho_D + \rho_m + \frac{\rho_D^2}{\rho_D+\rho_m}\right)$ | [48] |
| $Q_7$ | $3Hb\left(\rho_D + \rho_m + \frac{\rho_m^2}{\rho_D+\rho_m}\right)$ | [48] |

In what follows, we implement the above interaction term in the context of HRDE and EHRDE and using observational data to obtain the best values of each model's parameters. We also survey the cosmographical aspects of the models and then find the best models among the other ones using the Bayesian Evidence method. Finally, we compare the models with ΛCDM as the reference model by the means of the modified CAMB code package, as well.

## 3 Holographic Ricci dark energy models

In this section we study the behavior of two most used types of holographic dark energy model namely, interacting Holographic Ricci dark energy (HRDE) and interacting extended holographic Ricci dark energy (EHRDE). First, we produce two coupled differential equations to be solved numerically. This coupled differential equation shows the behavior of dark energy and Hubble parameter, suitable for both interacting and non-interacting models rather than using an analytical solution for them. Secondly, we find the cosmological parameters of each case and provide the results of the analysis in relevant tables.

### 3.1 Interacting holographic Ricci dark energy model

In a spatially flat Universe, the holographic dark energy is proportional to Ricci scalar curvature [38]

$$R = -6\left(\dot{H} + 2H^2\right),\tag{10}$$

and it is well-known that the density of dark energy can be written as [38]

$$\rho_D = 3\alpha M_P^2 \left(\dot{H} + 2H^2\right),\tag{11}$$

in which $H = \frac{\dot{a}}{a}$ is the Hubble parameter denoting the expansion rate of the Universe, the dot denotes the derivative in terms of $t$, $\alpha$ is a dimensionless parameter should be constrained as a free parameter and $M_P = 1/\sqrt{8\pi G}$ is the reduced Planck mass and $G$ is the Newton constant. Taking time derivative of Eq. (4) and using Eqs. (4), (8), (10) and (11) one can obtain the following coupled differential equations

$$\Omega_D' = \left(2\left(1 - \Omega_D\right)\left(\frac{\Omega_D}{\alpha} - 2\right) + 3\left(1 - \Omega_D - \Omega_i\right)\right),\tag{12}$$

$$H' = H\left(\frac{\Omega_D}{\alpha} - 2\right),\tag{13}$$

and

$$\Omega_i = \frac{Q}{3M_P^2 H^3}.\tag{14}$$

in which $\dot{\Omega}_D = \Omega_D' H$ and $\dot{H} = H'H$ where the prime denotes derivative with respect to $x = \ln a$ and $a = (1+z)^{-1}$. Then, the evolution of the density of dark energy and the Hubble parameter for HRDE in terms of redshift, after some algebra can be written as

$$\frac{d\Omega_D}{dz} = -\left(\frac{1}{1+z}\right)\left(2\left(1 - \Omega_D\right)\left(\frac{\Omega_D}{\alpha} - 2\right)\right.$$
$$\left. + 3\left(1 - \Omega_D - \Omega_i\right)\right),\tag{15}$$

$$\frac{dH}{dz} = -\left(\frac{H}{1+z}\right)\left(\frac{\Omega_D}{\alpha} - 2\right).\tag{16}$$

The results of the numerical calculation of these two coupled equations according to the combined observational data (see Appendix A for more details) can be seen in the Tables 2 and 3.

### 3.2 Interacting extended holographic Ricci dark energy model

A flexible form of HRDE has been proposed as the Extended HRDE with the following form of the dark energy density [39]

$$\rho_D = 3M_P^2 \left(\beta\dot{H} + \alpha H^2\right),\tag{17}$$

where $\alpha$ and $\beta$ are constant parameters to be constrained by observational data, $M_P = 1/\sqrt{8\pi G}$ is the reduced Planck mass and $G$ is the Newton constant. It is clear that by the assumption of $\beta = \alpha_{ERDE}/2 = \alpha_{RDE}$ the EHRDE reduces to HRDE. Again, taking time derivative of Eq. (4) and using Eqs. (4), (8), (17) and $\Omega' = \frac{\dot{\Omega}}{H}$ it is possible to reach the following coupled differential equations

**Table 2** The fitted values of cosmological parameters for the holographic Ricci dark energy model (Eqs. (15) and (16)) using linear and non-linear interactions in Table 1. The Pantheon supernova data, BAO BOSS DR12, CMB Planck 2015, fgas( gas mass fraction) and SZ/Xray (Sunyaev–Zeldovich effect and X-ray emission) data from galaxy clusters (GC) data has been used (See Appendix A)

| | Linear interactions | | | |
|---|---|---|---|---|
| Params | N/A | $3Hb\rho_D$ | $3Hb\rho_m$ | $3Hb\left(\rho_D+\rho_m\right)$ |
| $H_0$ | $68.8878^{+0.5712}_{-0.5668}$ | $68.972^{+0.5961}_{-0.5987}$ | $68.8885^{+0.5845}_{-0.5821}$ | $68.9558^{+0.6072}_{-0.6066}$ |
| $\Omega_D$ | $0.7220^{+0.01787}_{-0.01535}$ | $0.6965^{+0.0131}_{-0.0162}$ | $0.7053^{+0.01589}_{-0.01511}$ | $0.6813^{+0.0131}_{-0.0200}$ |
| $\alpha$ | $0.4399^{+0.0489}_{-0.0465}$ | $0.4240^{+0.0457}_{-0.0561}$ | $0.4296^{+0.0521}_{-0.0520}$ | $0.4150^{+0.0412}_{-0.0601}$ |
| $b$ | – | $0.0378^{+0.0209}_{-0.0207}$ | $0.0376^{+0.0052}_{-0.0052}$ | $0.0343^{+0.0046}_{-0.0046}$ |
| $M$ | $-19.3867^{+0.0206}_{-0.0209}$ | $-19.3846^{+0.0205}_{-0.0207}$ | $-19.3864^{+0.0877}_{-0.0857}$ | $-19.3841^{+0.0211}_{-0.0221}$ |
| $b_{fgas}$ | $0.7685^{+0.0754}_{-0.0887}$ | $0.844^{+0.0156}_{-0.0155}$ | $0.8169^{+0.0154}_{-0.0155}$ | $0.8186^{+0.0142}_{-0.0162}$ |
| Age | $13.6126^{+0.4554}_{-0.4523}$ | $13.67307^{+0.3126}_{-0.3120}$ | $13.8396^{+0.4123}_{-0.4112}$ | $13.8601^{+0.3411}_{-0.5925}$ |
| $z_t$ | $0.5376^{+0.0978}_{-0.0678}$ | $0.5041^{+0.0701}_{-0.0702}$ | $0.5521^{+0.0733}_{-0.0733}$ | $0.5492^{+0.0867}_{-0.0864}$ |
| $\chi^2$ | 118.6778 | 117.3842 | 117.6833 | 117.7193 |
| $\chi_{dof}$ | 1.0231 | 1.0231 | 1.0145 | 1.0148 |

**Table 3** The fitted values of cosmological parameters for the holographic Ricci dark energy model (Eqs. (15) and (16)) using non-linear interactions listed in Table 1. The Pantheon supernova data, BAO BOSS DR12, CMB Planck 2015, fgas (gas mass fraction) and SZ/Xray (Sunyaev–Zeldovich effect and X-ray emission) data from galaxy clusters (GC) data has been used (see Appendix A)

| | Non-linear interactions | | | |
|---|---|---|---|---|
| Params | $3Hb\left(\rho_D+\frac{\rho_D^2}{\rho_D+\rho_m}\right)$ | $3Hb\left(\rho_m+\frac{\rho_m^2}{\rho_D+\rho_m}\right)$ | $3Hb\left(\rho_m+\rho_D+\frac{\rho_D^2}{\rho_D+\rho_m}\right)$ | $3Hb\left(\rho_m+\rho_D+\frac{\rho_m^2}{\rho_D+\rho_m}\right)$ |
| $H_0$ | $68.8035^{+0.5877}_{-0.5829}$ | $68.8097^{+0.5875}_{-0.5761}$ | $68.907^{+05722}_{-0.5761}$ | $68.8798^{+0.5976}_{-0.5978}$ |
| $\Omega_D$ | $0.6911^{+0.0121}_{-0.0161}$ | $0.7039^{+0.0142}_{-0.0161}$ | $0.6777^{+0.0122}_{-0.0193}$ | $0.6826^{+0.0122}_{-0.0191}$ |
| $\alpha$ | $0.4257^{+0.0339}_{-0.0587}$ | $0.4302^{+0.0371}_{-0.0597}$ | $0.4157^{+0.0180}_{-0.0392}$ | $0.4182^{+0.0231}_{-0.0501}$ |
| $b$ | $0.0306^{+0.0047}_{-0.0047}$ | $0.0342^{+0.0051}_{-0.0052}$ | $0.0303^{+0.0052}_{-0.0052}$ | $0.0312^{+0.0052}_{-0.0052}$ |
| $M$ | $-19.3865^{+0.0212}_{-0.0211}$ | $-19.3887^{+0.0202}_{-0.0211}$ | $-19.3846^{+0.0200}_{-0.0201}$ | $-19.386^{+0.0222}_{-0.0211}$ |
| $b_{fgas}$ | $0.8561^{+0.0574}_{-0.0663}$ | $0.8186^{+0.0587}_{-0.0515}$ | $0.8971^{+0.0142}_{-0.0142}$ | $0.8810^{+0.0154}_{-0.0155}$ |
| Age | $13.69078^{+0.3020}_{-0.5872}$ | $14.0109^{+0.3211}_{-0.3991}$ | $13.8764^{+0.3885}_{-0.7655}$ | $14.0221^{+0.3002}_{-0.5967}$ |
| $z_t$ | $0.5495^{+0.0870}_{-0.0781}$ | $0.5671^{+0.1102}_{-0.0955}$ | $0.5711^{+0.0677}_{-0.0896}$ | $0.5532^{+0.0911}_{-0.1401}$ |
| $\chi^2$ | 117.4945 | 117.7134 | 117.8415 | 117.7102 |
| $\chi_{dof}$ | 1.0129 | 1.0148 | 1.0159 | 1.0147 |

$$\Omega'_D = \left(2\left(1-\Omega_D\right)\left(\frac{\Omega_D}{2\beta}-\frac{2\alpha-3\beta}{2\beta}-\frac{3}{2}\right)\right.$$
$$\left.+\frac{2\alpha}{\beta}-\left(\frac{2\alpha-3\beta}{\beta}\right)-3\left(\Omega_D+\Omega_i\right)\right), \quad (18)$$

$$H' = H\left(\frac{\Omega_D}{2\beta}-\frac{2\alpha-3\beta}{2\beta}-\frac{3}{2}\right), \quad (19)$$

and

$$\Omega_i = \frac{Q}{3M_p^2 H^3}. \quad (20)$$

in which $\dot{\Omega}_D = \Omega'_D H$ and $\dot{H} = H'H$ where the prime denotes derivative with respect to $x = \ln a$ and $a = (1+z)^{-1}$. Then, the evolution of the density of dark energy and the Hubble parameter for EHRDE in terms of redshift can be written as

$$\frac{d\Omega}{dz} = -\left(\frac{1}{1+z}\right)\left(2\left(1-\Omega_D\right)\left(\frac{\Omega_D}{2\beta}-\frac{2\alpha-3\beta}{2\beta}-\frac{3}{2}\right)\right.$$
$$\left.+\frac{2\alpha}{\beta}-\left(\frac{2\alpha-3\beta}{\beta}\right)-3\left(\Omega_D+\Omega_i\right)\right), \quad (21)$$

$$\frac{dH}{dz} = -\left(\frac{H}{1+z}\right)\left(\frac{\Omega_D}{2\beta}-\frac{2\alpha-3\beta}{2\beta}-\frac{3}{2}\right). \quad (22)$$

The results of the numerical calculation of these two coupled equations according to the combined observational data (see Appendix A for more details) can be seen in the Tables 4 and 5.

## 4 Cosmography

For studying the cosmographical behavior of the model, we may check the behavior of the Hubble parameter Eq. (1), the deceleration parameter Eq. (2) and the jerk parameter Eq. (3). In addition to the calculation of the deceleration parameter, we obtain the transition redshift. For this, we employ the well-known Brent's method which uses the combination of some methods with inverse quadratic interpolation as a secured version of the secant algorithm. This method by using three prior points can estimate the zero crossing. A description of this method can be found in "Numerical Recipes in C" handbook [93].

**Table 4** The fitted values of cosmological parameters for the extended holographic Ricci dark energy model (Eqs. (21) and (22)) using linear and non-linear interactions in Table 1. The Pantheon supernova data, BAO BOSS DR12, CMB Planck 2015, fgas (gas mass fraction) and SZ/Xray (Sunyaev–Zeldovich effect and X-ray emission) data from galaxy clusters (GC) data has been used (see Appendix A)

| | Linear interactions | | | |
|---|---|---|---|---|
| Params | N/A | $3Hb\rho_D$ | $3Hb\rho_m$ | $3Hb(\rho_D + \rho_m)$ |
| $H_0$ | $68.9821^{+1.2332}_{-1.5601}$ | $68.9841^{+0.8535}_{-1.3011}$ | $68.8995^{+0.8721}_{-1.1150}$ | $68.9523^{+0.79}_{-1.4}$ |
| $\Omega_D$ | $0.6968^{+0.0142}_{-0.0175}$ | $0.6972^{+0.0071}_{-0.0103}$ | $0.7054^{+0.0101}_{-0.0088}$ | $0.6818^{+0.0120}_{-0.0121}$ |
| $\alpha$ | $0.4432^{+0.03001}_{-0.0889}$ | $0.4232^{+0.0441}_{-0.0733}$ | $0.4266^{+0.0221}_{-0.0377}$ | $0.4027^{+0.0321}_{-0.0551}$ |
| $\beta$ | $0.8497^{+0.3247}_{-0.2123}$ | $0.8477^{+0.1222}_{-0.1221}$ | $0.8565^{+0.0551}_{-0.1102}$ | $0.8226^{+0.0612}_{-0.1127}$ |
| $b$ | – | $0.0378^{+0.0171}_{-0.0050}$ | $0.0369^{+0.0082}_{-0.0022}$ | $0.0350^{+0.0131}_{-0.0070}$ |
| $M$ | $-19.3853^{+0.1097}_{-0.1002}$ | $-19.3846^{+0.0291}_{-0.0291}$ | $-19.3868^{+0.0299}_{-0.0291}$ | $-19.3851^{+0.0311}_{-0.0310}$ |
| $b_{fgas}$ | $0.8429^{+0.1175}_{-0.0998}$ | $0.8442^{+0.1120}_{-0.0587}$ | $0.8173^{+0.5022}_{-0.4874}$ | $0.8884^{+0.0411}_{-0.0361}$ |
| $Age$ | $13.4711^{+0.6564}_{-0.6303}$ | $13.6867^{+0.6721}_{-0.6721}$ | $13.8291^{+0.8991}_{-0.8991}$ | $13.8497^{+0.7211}_{-0.7211}$ |
| $z_t$ | $0.5411^{+0.1342}_{-0.05401}$ | $0.5456^{+0.1601}_{-0.0461}$ | $0.5473^{+0.0177}_{-0.0491}$ | $0.5412^{+0.1601}_{-0.0571}$ |
| $\chi^2$ | 118.0536 | 117.3597 | 117.6568 | 117.6867 |
| $\chi_{dof}$ | 1.0159 | 1.0117 | 1.0143 | 1.0145 |

**Table 5** The fitted values of cosmological parameters for the extended holographic Ricci dark energy model (Eqs. (21) and (22)) using linear and non-linear interactions listed in Table 1. The Pantheon supernova data, BAO BOSS DR12, CMB Planck 2015, fgas (gas mass fraction) and SZ/Xray (Sunyaev–Zeldovich effect and X-ray emission) data from galaxy clusters (GC) data has been used (see Appendix A)

| | Non-linear interactions | | | |
|---|---|---|---|---|
| Params | $3Hb\left(\rho_D + \frac{\rho_D^2}{\rho_D+\rho_m}\right)$ | $3Hb\left(\rho_m + \frac{\rho_m^2}{\rho_D+\rho_m}\right)$ | $3Hb\left(\rho_m + \rho_D + \frac{\rho_D^2}{\rho_D+\rho_m}\right)$ | $3Hb\left(\rho_m + \rho_D + \frac{\rho_m^2}{\rho_D+\rho_m}\right)$ |
| $H_0$ | $68.8156^{+0.7270}_{-0.8431}$ | $68.8289^{+0.8411}_{-1.1231}$ | $68.9547^{+0.7005}_{-0.7054}$ | $68.8765^{+0.8639}_{-1.4052}$ |
| $\Omega_D$ | $0.6914^{+0.0112}_{-0.0091}$ | $0.7043^{+0.0111}_{-0.0131}$ | $0.6803^{+0.0113}_{-0.0086}$ | $0.6833^{+0.0099}_{-0.0098}$ |
| $\alpha$ | $0.4238^{+0.0364}_{-0.0511}$ | $0.4316^{+0.0247}_{-0.0451}$ | $0.4024^{+0.0197}_{-0.0385}$ | $0.4171^{+0.0233}_{-0.0451}$ |
| $\beta$ | $0.8494^{+0.1001}_{-0.1001}$ | $0.8605^{+0.0433}_{-0.0925}$ | $0.8246^{+0.1444}_{-0.1224}$ | $0.8352^{+0.0365}_{-0.0891}$ |
| $b$ | $0.0314^{+0.0030}_{-0.01312}$ | $0.0335^{+0.01370}_{-0.0051}$ | $0.0325^{+0.0078}_{-0.0142}$ | $0.0315^{+0.0072}_{-0.0020}$ |
| $M$ | $-19.3874^{+0.0301}_{-0.0301}$ | $-19.3885^{+0.1038}_{-0.1168}$ | $-19.3851^{+0.0380}_{-0.0380}$ | $-19.3865^{+0.0410}_{-0.0410}$ |
| $b_{fgas}$ | $0.8556^{+0.1523}_{-0.1915}$ | $0.819^{+0.0655}_{-0.0900}$ | $0.8892^{+0.1331}_{-0.1602}$ | $0.8803^{+0.3391}_{-0.5254}$ |
| $Age$ | $13.7062^{+0.6012}_{-0.9110}$ | $14.0112^{+0.9010}_{-0.9010}$ | $13.9164^{+0.8440}_{-1.0204}$ | $14.0429^{+0.7972}_{-0.7980}$ |
| $z_t$ | $0.5563^{+0.0652}_{-0.0491}$ | $0.5632^{+0.1300}_{-0.0541}$ | $0.5701^{+0.1101}_{-0.0371}$ | $0.5743^{+0.1210}_{-0.0523}$ |
| $\chi^2$ | 117.4945 | 117.7134 | 117.7193 | 117.7102 |
| $\chi_{dof}$ | 1.0129 | 1.0148 | 1.0148 | 1.0147 |

The deceleration parameter of two interacting models of Ricci dark energy by substitution of the Eqs. (16), (22) into Eq. (2) can be written as

$$q_{HRDE} = -1 - \left(\frac{\Omega_D}{\alpha} - 2\right), \tag{23}$$

**Table 6** The definition of strength of evidence for model $M_i$ according to Jeffrey's scale [121]

| Value of evidence | Strength of evidence |
|---|---|
| $\Delta \ln B_{ij} < 1$ | Insignificant |
| $1 < \Delta \ln B_{ij} < 3$ | Substantial |
| $3 < \Delta \ln B_{ij} < 4$ | Strong |
| $\Delta \ln B_{ij} \geqslant 5$ | Very strong |

$$q_{EHRDE} = -1 - \left(\frac{\Omega_D}{2\beta}\right.$$
$$\left.\frac{2\alpha - 3\beta}{2\beta} - \frac{3}{2}\right), \tag{24}$$

 

respectively. Moreover, the value of $q_0$ for all models according to the best values of fitted parameters presented in Tables 2, 3, 4, 5 is approximately around $q_0 \approx -0.63$ for the present time which has good agreement with the value of the deceleration parameter by Planck ($q_0 = -0.55$) [90] and shows an accelerated expansion of the Universe. The behavior of the deceleration parameter versus redshift for all models is plotted in Figs. 1 and 2. According to the plotted results in Figs. 1 and 2, one can see that in region of $1\sigma$ interval level both lower and upper bounds of interacting and non-interacting HRDE and EHRDE models show the accelerating expansion within the redshift range $z = (0.4, 0.8)$ which compatible with the recent observational works [81,94–97] (to mention few). The obtained results in the Tables 2, 3, 4 and 5 by the use of observational data show that the values of the transition redshift ($z_t$) of all models is in range of recent obtained values for transition redshift $z_t = [0.4, 1]$ [81,94–97] (to mention few).

For the jerk parameter by substitution of Eqs. (23) and (24) into Eq. (3) and with help of Eqs. (11) and (17), after some algebra, we may obtain

$$j_{HRDE} = q_{HRDE} + 2q_{RDE}^2 + 2\left((1 - \Omega_D) \times \left(\frac{\Omega_D}{\alpha} - 2\right) + 3 - 3\Omega_D - 3\Omega_i\right), \quad (25)$$

$$j_{EHRDE} = q_{EHRDE} + 2q_{EHRDE}^2 + (2(1 - \Omega_D) \times \left(\frac{\Omega_D}{2\beta} - \frac{2\alpha - 3\beta}{2\beta} - \frac{3}{2}\right) + \frac{2\alpha}{\beta} - \left(\frac{2\alpha - 3\beta}{\beta}\right) - 3(\Omega_D + \Omega_i)). \quad (26)$$

According to the jerk parameter, we can explain the behavior of models in comparison to $\Lambda$CDM [85,98,99]. Compared to the deceleration parameter, the positive value for the jerk parameter demonstrates an accelerated expansion of the Universe. For $\Lambda$CDM in a flat Universe the value of jerk parameter has a constant tendency to $j = 1$ [85,98,99]. The observational constraints on the value of the cosmic jerk parameter in comparison with the deceleration parameter are weak $-5 < j_0 < 10$ [100–103]. In this work, we obtained the value of jerk parameters in a range of $1 < j_0 < 2$ and its behavior for the interacting HRDE and interacting EHRDE is plotted in Figs. 3 and 4, respectively. The value of the jerk parameter for interacting HRDE model remains positive and close to 1 between the redshift $z = [0.2, 0.6]$ and it crosses this line within the range of $1\sigma$. The EHRDE model has a tendency to reach the 1 at the early time and the values of its trajectory embrace the value of 1 in all redshifts within the range of $1\sigma$.

## 5 Observational analysis

In this section, we summarize the method used to analyze the models. In order to analyze the models, we used SNIa, BAO, CMB, SZ/Xray (Sunyaev–Zeldovich effect and X-ray emission) and fgas (gas mass fraction) data introduced in Appendix A. For this purpose we employed the public codes EMCEE [104] and GetDist Python package[1] for implementing the MCMC method and plotting the contours respectively.

### 5.1 Parameters

By minimizing the $\chi^2$ we may obtain the best values of cosmological parameters

$$\chi_{total}^2 = \chi_{Pantheon}^2 + \chi_{BAO}^2 + \chi_{CMB}^2 + \chi_{SZ/Xray}^2 + \chi_{Fgas}^2. \quad (27)$$

According to the obtained results listed in the Tables 2, 3, 4 and 5 we compare their compatibility with the very latest obtained cosmological parameters.

The Hubble constant $H_0$ is an important quantity in cosmology for calculation of age and size of the Universe and also is a key factor for measuring the brightness and the mass of stars. This quantity corresponds to the Hubble parameter at the time of observation. Using the observational data in this work we obtained the value of the Hubble parameter for all models and we found a good consistency with the latest observational data, $H_0 = 67.78_{-0.87}^{+0.91}$ [105], $H_0 = 68_{-4.1}^{+4.2}$ [106], $H_0 = 67.66_{-0.42}^{+0.42}$ [90] and $H_0 = 70_{-8}^{+12}$ [107]. The value of dark energy density $\Omega_D$ also for all models has suitable compatibility with the latest obtained value $\Omega_D = 0.692_{-0.012}^{+0.012}$ [90]. However, the $Q_4$ has the closest value between the studied models with $\Omega_D = 0.6911_{-0.0161}^{+0.0121}$ and $\Omega_D = 0.6914_{-0.0091}^{+0.0112}$ for HRDE and EHRDE respectively. For the HRDE model, in spite of employing the latest observational data the value of $\alpha$ has not faced with remarkable change compared to the previous works [78,108,109]. For further information, in the case of $\alpha_{HRDE} = \alpha_{EHRDE}$ and $\beta = 2\alpha_{EHRDE}$ the EHRDE model reduces to HRDE. This ratio for all models stays in the range of $\beta/\alpha_{EHRDE} = 2_{-0.01}^{+0.01}$. According to this definition and considering the best fit values listed in Tables 2, 3, 4 and 5 we can see that the EHRDE model has a strong tendency towards HRDE model.

The value of the depletion component of the bias factor related to the gas dynamical simulation from fgas (gas mass fraction) data has been obtained $b_{fgas} = 0.824_{-0.033}^{+0.033}$ [110]. This value has been used in some works as a fixed value [108, 111,112]. We found that fixing this parameter strongly affects the values of other parameters and the value of $\chi^2$ as well. For example by using $b_{fgas} = 0.824$ we obtained $\Omega_D = 0.66$ and $b = 0.08$. Taking the $b_{fgas}$ as a free parameter, we reached the bigger value for this quantity compared to Ref. [110] except for $Q_3$ and $Q_5$ having a smaller number. It should be mentioned that fixing this value makes the Universe older and out of the acceptable range of age. After fitting this value, the Age of the Universe for all models except for $Q_5$ and $Q_7$ is also in good agreement with the recent observational data ($Age_{Planck} = 13.79$Gyr) [90].

---

[1] https://getdist.readthedocs.io.

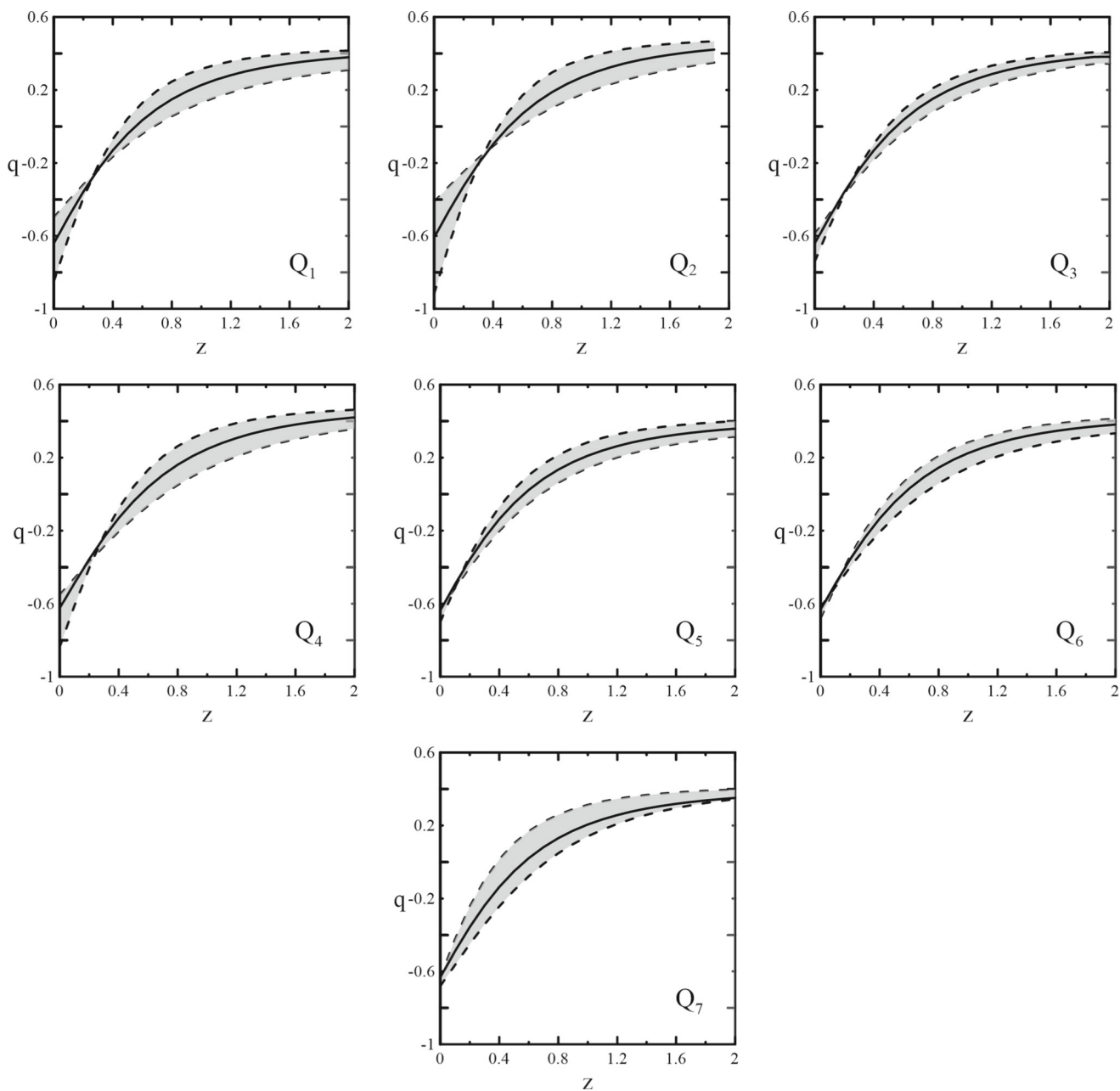

**Fig. 1** The evolution of the deceleration parameter in terms of redshift for HRDE model (Eqs. (15) and (16)) with the corresponding $1\sigma$ interval level according to the best fitted value listed in Tables 2 and 3 and using Eq. (23) presenting the deceleration parameter with interactions listed in Table 1

To compare the success of the models on fitting data we calculate $\chi_{dof}$ with $N = 116$ represents the entire data points used in this work. We notice that all the models are successful with the reasonable value of the goodness of freedom (DOF). The DOF value for $\Lambda$CDM model with $\chi^2 = 116.0528$ is $\chi_{dof} = 1.0004$ which is slightly better than the other models. The non-interacting HRDE and EHRDE models have the biggest values of the degree of freedom and $Q_2$ for both

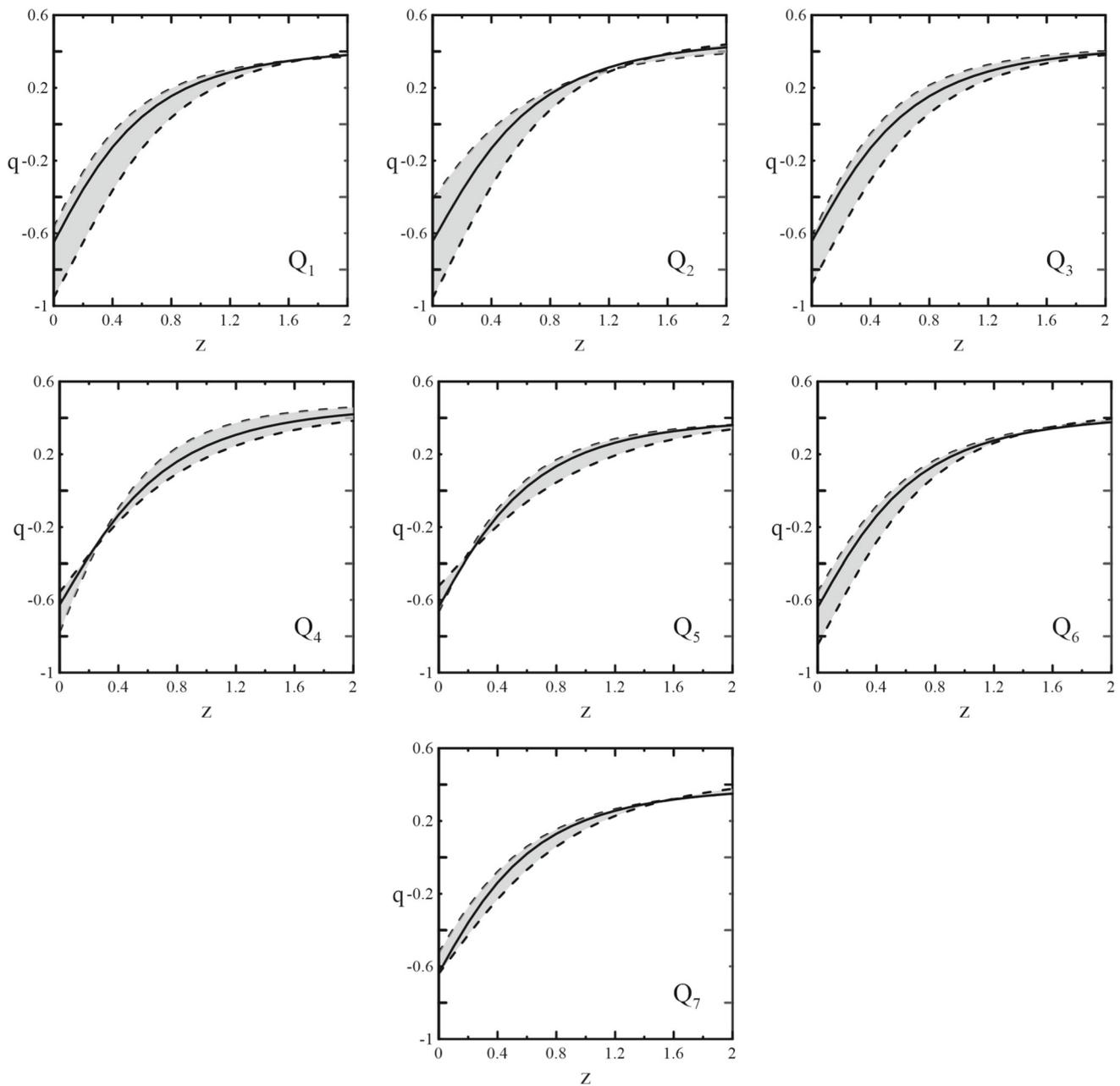

**Fig. 2** The evolution of the deceleration parameter in terms of redshift for EHRDE model (Eqs. (21) and (22)) with the corresponding $1\sigma$ interval level according to the best fitted value listed in Tables 4 and 5 and using Eq. (24) presenting the deceleration parameter with interactions listed in Table 1

HRDE and EHRDE shows the highest success on the fitting data.

These results show the consistency of HRDE and EHRDE models with the latest observational data and also considering the interaction between dark sectors (all types of interaction in this work) does not impose any problem on this issue. In addition, these results show that the fgas (gas mass fraction) and SZ/X-Ray data can play a rational role in the determination of free parameters for the cosmological models.

Clearly, for both interacting and non-interacting Ricci dark energy model the values of $\chi^2$ in case of existence of interaction are smaller than the non-interacting models which are due to the additional parameter $b$. The interacting and non-interacting models have a bigger value of $\chi^2$ compared to the $\Lambda$CDM. From these analyses also it can be shown that the linear interaction terms lead to a bigger value of the decoupling constant ($b$) in comparison with phenomenological interactions ( see Tables 2, 3, 4 and 5).

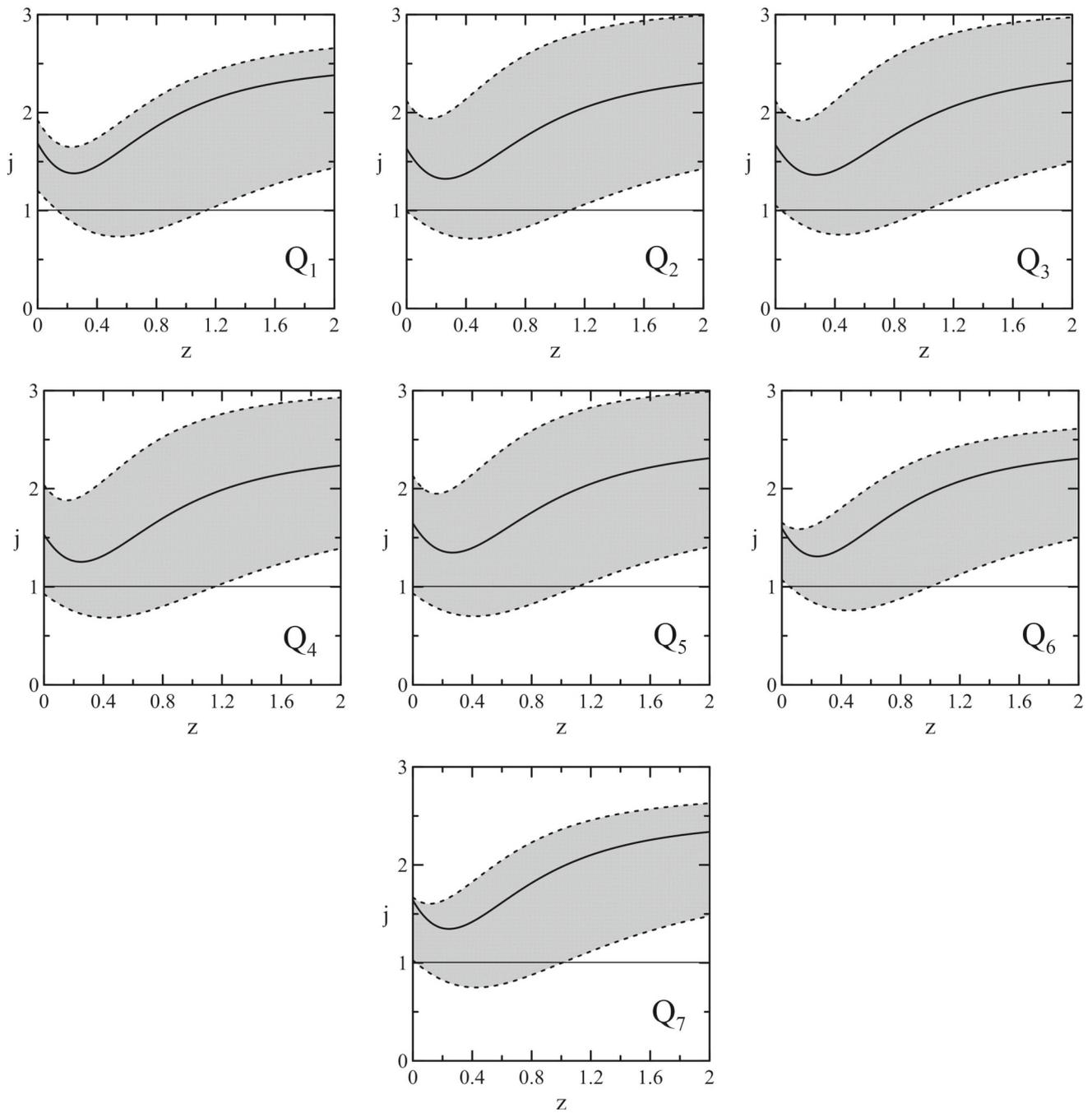

**Fig. 3** The evolution of the jerk parameter in terms of redshift for HRDE model (Eqs. (15) and (16)) with the corresponding $1\sigma$ interval level according to the best fitted value listed in Tables 2 and 3 and using Eq. (25) presenting the jerk parameter with interactions listed in Table 1. The straight line denotes the $\Lambda$CDM

### 5.2 Bayesian evidence

To determine the best cosmological models among several studied models we cannot rely on the fitted values of the relevant parameters. Even though minimizing $\chi^2$ is the most simple way to get the best fitting of free parameters, it is usually unreasonable to distinguish the best model between a variety of studied models. Hence, for this issue Akaike Information Criterion (AIC) [113] and Bayesian Information Criterion (BIC) [114] have been proposed. For additional information see [115–118]. The AIC model selection function can be expressed as

$$AIC = -2\ln \mathcal{L}_{max} + 2k, \qquad (28)$$

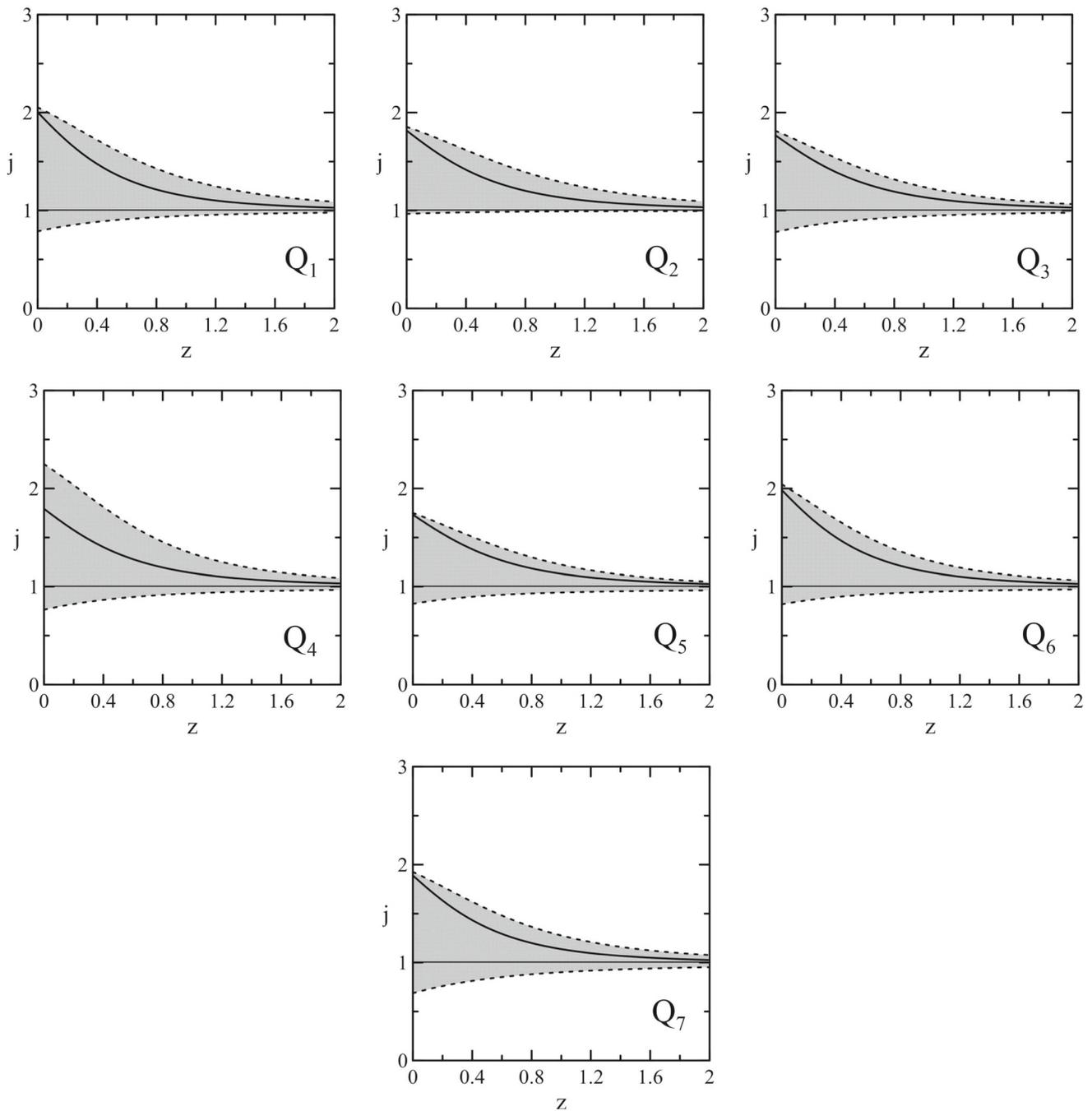

**Fig. 4** The evolution of the jerk parameter in terms of redshift for EHRDE model (Eqs. (21) and (22)) with the corresponding $1\sigma$ interval level according to the best fitted value listed in Tables 4 and 5 and using Eq. (26) presenting the jerk parameter with interactions listed in Table 1. The straight line denotes the $\Lambda$CDM

where $-2\ln\mathcal{L}_{max} = \chi^2_{min}$ is the highest likelihood, $k$ is the number of free parameters and N is the number of data points used in the analysis. The BIC is similar to AIC with the different second term

$$BIC = -2\ln\mathcal{L}_{max} + k\ln N. \tag{29}$$

It is obvious that a model favored by observations should give a small AIC and a small BIC.

The level of support for each model from AIC is

- **Less than 2**: This indicates there is substantial evidence to support the model (i.e., the model can be considered almost as good as the best model).

- **Between 4 and 7**: This indicates that the model has considerably less support.
- **Between 8 and 10 or bigger**: This indicates that there is essentially no support for the model (i.e., it is unlikely to be the best model).

The level of the evidence against models if the tool of selection is BIC:

- **Less than 2**: It is not worth more than a bare mention (i.e., the model can be considered almost as good as the best model).
- **Between 2 and 6**: The evidence against the model is positive.
- **Between 6 and 10**: The evidence against the candidate model is strong. (i.e., it is unlikely the best model).
- **Bigger than 10**: The evidence is very strong (i.e., it is unlikely to be the best model).

Furthermore, It is more capable to check the models using a more complicated method so-called Bayesian Evidence (BE). Originating from Bayes theorem stating that the posterior probability of the parameter $\theta$ can be defined as

$$P\left(\theta \mid D, M\right) = \frac{P\left(D \mid \theta, M\right) P\left(\theta \mid M\right)}{P\left(D \mid M\right)} \tag{30}$$

and according to the definition of the conditional probability one may express posterior probability of a model as

$$BE = P\left(D \mid M\right) = \int_{\theta} P\left(D \mid \theta, M\right)$$
$$P\left(\theta \mid M\right) d^{n}\theta \tag{31}$$

in which we take an integral over the entire parameter space of the likelihood. The $P\left(x \mid \theta, M\right)$ is the likelihood, $P\left(\theta \mid M\right)$ is the prior information, $\theta$ denotes the parameters of the model and $n$ is the number of parameters making a $n$ dimensional integration. Taking an explicit integration over the parameter space for high dimensional models (Here 5 and 6 for HRDE and EHRDE respectively) is a very time consuming and expensive task. Therefore we use the nested sampling to decrease the time of calculation significantly. For this issue, we use the Dynamic Nested Sampling (dynesty)[2] publicly available Python package for estimating the evidences [119].

For any two models (under investigation model $M_i$ and the reference model $M_j$ (Here $\Lambda$CDM)) the following equation holds as Bayes factor conveying the way of supporting the model $M_i$ over $M_j$ by observational data [120].

$$B_{ij} = \frac{P\left(D \mid M_i\right)}{P\left(D \mid M_j\right)} \tag{32}$$

---

[2] https://dynesty.readthedocs.io.

**Table 7** The strength of Jefrey's scale $\ln B_{ij}$ for holographic RIcci dark energy model (HRDE) compared to $\Lambda$CDM model. The negative sign demonstrates that the $\Lambda$CDM is superior to HRDE model

| Model | $\Delta \ln B_{ij}$ |
|---|---|
| $N/A$ | $-1.7861 \pm 0.024$ |
| $3Hb\rho_D$ | $-2.0454 \pm 0.031$ |
| $3Hb\rho_m$ | $-1.9845 \pm 0.031$ |
| $3Hb(\rho_D + \rho_m)$ | $-2.1778 \pm 0.032$ |
| $3Hb(\rho_D + \frac{\rho_D^2}{\rho_D + \rho_m})$ | $-2.3549 \pm 0.031$ |
| $3Hb(\rho_m + \frac{\rho_m^2}{\rho_D + \rho_m})$ | $-2.1682 \pm 0.029$ |
| $3Hb(\rho_D + \rho_m + \frac{\rho_D^2}{\rho_D + \rho_m})$ | $-2.3965 \pm 0.030$ |
| $3Hb(\rho_D + \rho_m + \frac{\rho_m^2}{\rho_D + \rho_m})$ | $-2.2037 \pm 0.028$ |

**Table 8** The strength of Jefrey's scale $\ln B_{ij}$ for holographic RIcci dark energy model (EHRDE) compared to $\Lambda$CDM model. The negative sign demonstrates that the $\Lambda$CDM is superior to EHRDE model

| Model | $\Delta \ln B_{ij}$ |
|---|---|
| $N/A$ | $-1.8602 \pm 0.019$ |
| $3Hb\rho_D$ | $-2.3756 \pm 0.025$ |
| $3Hb\rho_m$ | $-2.2350 \pm 0.025$ |
| $3Hb(\rho_D + \rho_m)$ | $-2.5235 \pm 0.026$ |
| $3Hb(\rho_D + \frac{\rho_D^2}{\rho_D + \rho_m})$ | $-2.7110 \pm 0.023$ |
| $3Hb(\rho_m + \frac{\rho_m^2}{\rho_D + \rho_m})$ | $-2.4305 \pm 0.022$ |
| $3Hb(\rho_D + \rho_m + \frac{\rho_D^2}{\rho_D + \rho_m})$ | $-2.7423 \pm 0.023$ |
| $3Hb(\rho_D + \rho_m + \frac{\rho_m^2}{\rho_D + \rho_m})$ | $-2.5760 \pm 0.021$ |

Calculating the Bayes factor for each model and using Jefrey's scale [120,121] we compare the models. In Table 6 we provide the condition of supporting the models using Jefrey's scale and also listed the obtained values of this quantity for each model in the Tables 7 and 8. The negative value of $\Delta \ln B_{ij}$ denotes that the $\Lambda$CDM model is preferred over the RDE and ERDE models (Figs. 5, 6).

According to the results of Bayesian evidences shown in Tables 7 and 8, one can decide about choosing the appropriate interaction term. We take the $\Lambda$CDM model ($H_0 = 68.5846^{+0.7970}_{-0.8015}$, $\Omega_D = 0.6968^{+0.0174}_{-0.0173}$, $M = -19.3868^{+0.0202}_{-0.0207}$ and $bfg = 0.8280^{+0.0366}_{-0.0373}$) as the reference model for making comparison between models and obviously the measured values of $\Delta \ln B_{ij} = \ln B_i - \ln B_j$ are respect to the reference model. It is clear that the different types of Ricci dark energy model is not favored by observational data. But it should be noted that the proposing of holographic dark energy models is a way to overcome the problems with which $\Lambda$CDM is faced [21–24,30,31,40–43]. In spite of very small differences between the val-ues of $\Delta \ln B_{ij}$ of the models, according to the analysis above for two Ricci models namely, holographic Ricci dark energy model (HRDE) and extended holographic Ricci dark energy model (EHRDE) the best interaction models are $Q_3 = 3Hb\rho_m$ in the linear interaction's category and $Q_5 = 3Hb\left(\rho_m + \frac{\rho_m^2}{\rho_D + \rho_m}\right)$ in the non-linear interaction's category. The constraints on free parameters are summarized in Figs. 7, 8, 9, 10 and Tables 2, 3, 4 and 5. By adding the fgas (gas mass fraction) and SZ/X-Ray (Sunyaev–Zeldovich effect and X-ray emission) data compared to the previous works [78–80], we found that the best interaction is the linear one $Q_3 = 3Hb\rho_m$. In conclusion, in both models the interaction terms $Q_2$, $Q_3$ and $Q_5$ show better results in comparison with other interaction terms.

**Fig. 5** The CMB temperature spectra $c_\ell^{TT}$ of the HRDE (see Eqs. (15), (16)) with interaction terms listed in Table 1. All models' $\Omega_D$ and $H_0 \equiv 100h[\mathrm{km/s \cdot Mpc}]$ parameter are set as their best fitting values from Table 2 and 3. All models have equal $\Omega_b h^2 = 0.022$, $\Omega_\nu h^2 = 0.00064$ and manually tuned $\Omega_{cdm}$ so that $\Omega_b + \Omega_\nu + \Omega_{cdm} = 1 - \Omega_D$. Relative to $\Lambda$CDM model ($H_0 = 68.5846^{+0.7970}_{-0.8015}$, $\Omega_D = 0.6968^{+0.0174}_{-0.0173}$), the interacting and non-interacting HRDE models have the trend of yielding equal degree of anisotropies at larger angle scale or small $\ell$-poles

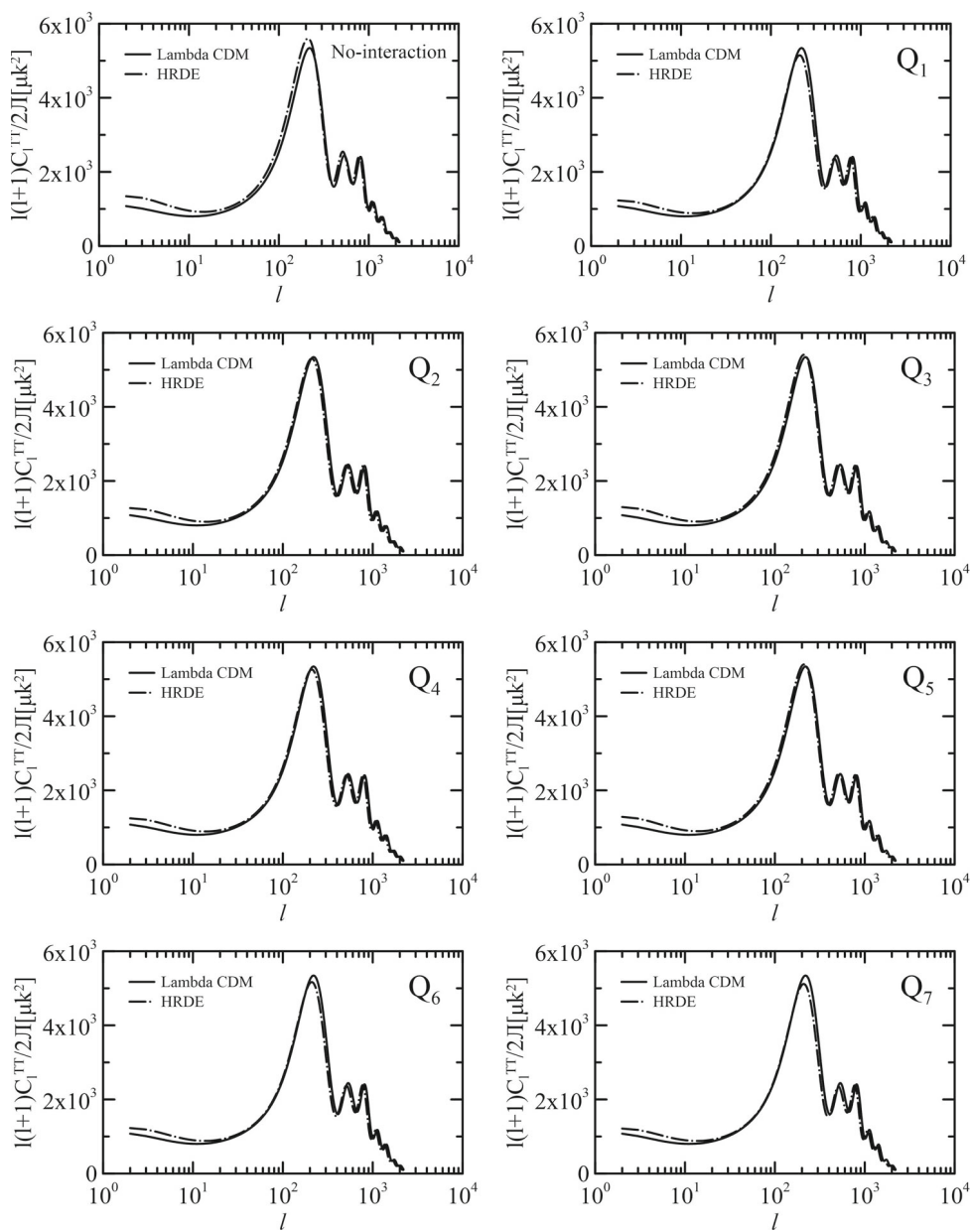

# 6 CMB power spectrums

In this section by the use of the modified version of the Boltzmann code CAMB[3] [122,123], we calculate and compare the power spectrums of the cosmic microwave anisotropy in all interacting and non-interacting HRDE and EHRDE models. We solve the Eqs. (15) and (16) for RDE model and Eqs. (21) and (22) for ERDE model to find out the effective equation of state coefficient $\omega(z) \equiv \frac{p_{DE}}{\rho_{DE}}$ and substitute it into CAMB to get the power spectrum. We do not consider perturbations of the dark energy. Our results of the temperature power spectrum ($TT$) according to the fitted results are depicted in Figs. 5 and 6. From the figures, we see that both the HRDE (Eqs. (15), (16)) and the EHRDE (Eqs. (21), (22)) models for all types of interactions show the trends of squeezing power spectrum of the cosmic microwave anisotropy to small $\ell$ or large angle scales. This squeezing can also be seen from the power spectrum of the matter distribution in the Universe. Embodying on the large scale structure of matter distributions, all these models exhibit an as high as 20% peak power spectrum's suppressing which occurs in small $k$ or large scale region. The origin of this suppression is mainly due to the relative lower baryon $\Omega_b = 0.0464$ and neutrino $\Omega_\nu = 0.00134$ occupation fraction in the cosmic contents partition scheme relative to $\Lambda$CDM model which are $\Omega_b = 0.0468$ and neutrino $\Omega_\nu = 0.00136$ while the lower (relative to $\Lambda$CDM) $\Omega_b$ and $\Omega_\nu$ values originate from our fitting results of smaller $H_0$ and larger $\Omega_D$ from SNIa, BAO and SZ/Xray data in the previous section and fixing choice of $\Omega_b h^2 = 0.022$, $\Omega_\nu h^2 = 0.00064$. Using $\Lambda$CDM as the reference model, we may observe that, $Q_2 = 3 H b \rho_D$ and $Q_3 = 3 H b \rho_m$ are the interaction models which are most close to $\Lambda$CDM. On the other hand, except for some very special cases, the EHRDE model as a whole does not exhibit manifest advantage over the simple HRDE model.

---

[3] https://camb.info.

**Fig. 6** The CMB temperature spectra $c_\ell^{TT}$ of the EHRDE (see Eqs. (21), (22)) with interaction terms listed in Table 1. All models' $\Omega_D$ and $H_0 \equiv 100 h [\text{km/s} \cdot \text{Mpc}]$ parameter are set as their best fitting values from Table 4 and 5. All models have equal $\Omega_b h^2 = 0.022$, $\Omega_\nu h^2 = 0.00064$ and manually tuned $\Omega_{\text{cdm}}$ so that $\Omega_b + \Omega_\nu + \Omega_{\text{cdm}} = 1 - \Omega_D$. Relative to $\Lambda$CDM model ($H_0 = 68.5846^{+0.7970}_{-0.8015}$, $\Omega_D = 0.6968^{+0.0174}_{-0.0173}$), the interacting and non-interacting HERDE models have the trend of yielding equal degree of anisotropies at larger angle scale or small $\ell$-poles

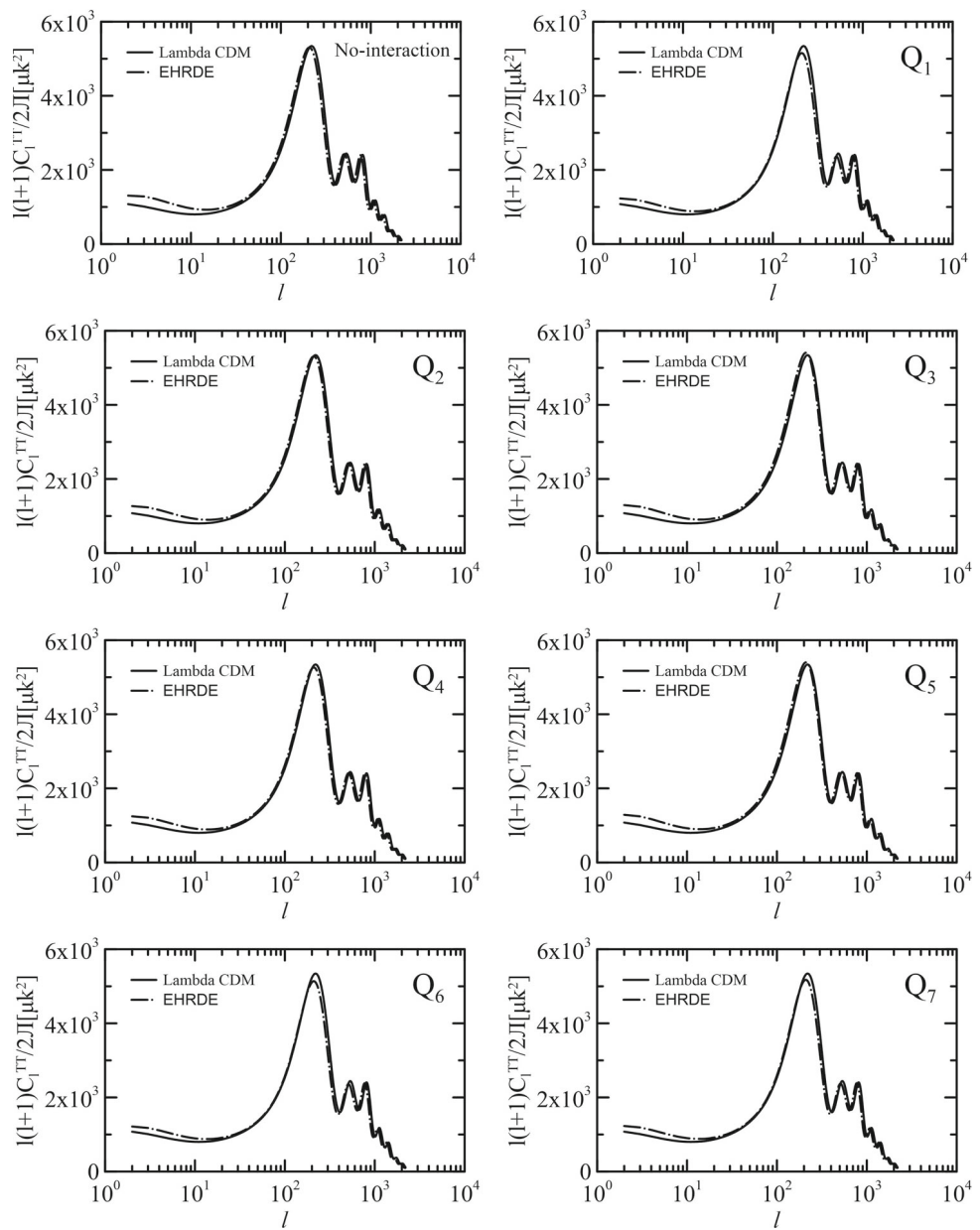

## 7 Conclusion

In this work, we compared the behavior of seven types of interaction case ($Q_1 = 3 H b (\rho_D + \rho_m)$, $Q_2 = 3 H b \rho_D$, $Q_3 = 3 H b \rho_m$, $Q_4 = 3 H b \left( \rho_D + \frac{\rho_D^2}{\rho_D + \rho_m} \right)$, $Q_5 =$

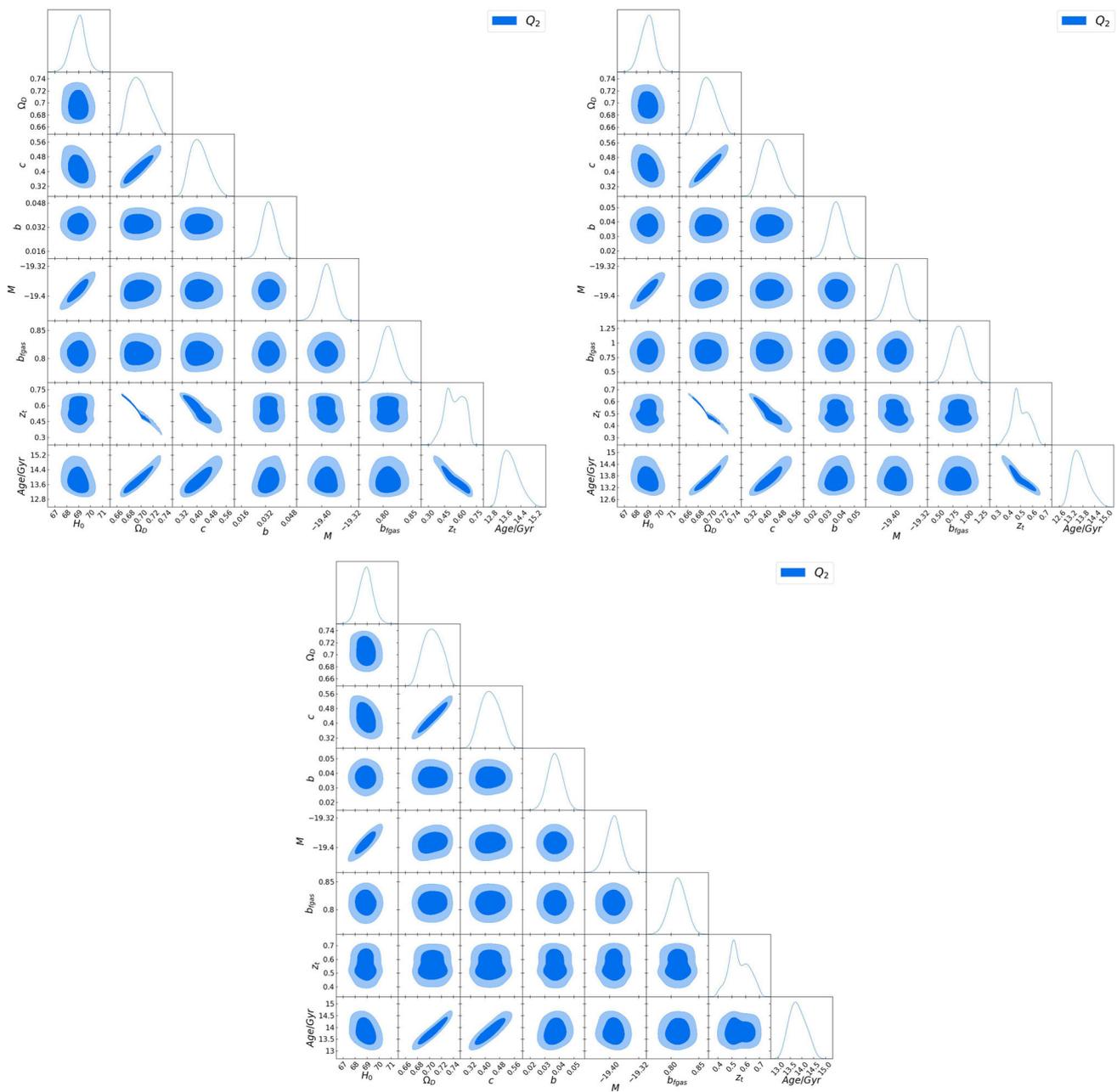

**Fig. 7** The contour maps of the HRDE (see Eqs. (15) and (16)) with three types of linear interaction $Q_1$, $Q_2$ and $Q_3$ listed in Table 1. In this figure $H_0$ is the Hubble parameter, $\Omega_D$ is the dark energy density, $\alpha = c^2$ is the dimensionless parameter, $b$ is the coupling constant, $M$ is the nuisance parameter of SNIa data, $b_{fgas}$ is the nuisance parameter of fgas mass fraction data, $z_t$ is the transition redshift and $Age$ is the age of the Universe for the HRDE model. The best-fitted values of these parameters are listed in Table 2

$3Hb\left(\rho_m + \frac{\rho_m^2}{\rho_D + \rho_m}\right)$, $Q_6 = 3Hb\left(\rho_D + \rho_m + \frac{\rho_D^2}{\rho_D + \rho_m}\right)$ and $Q_7 = 3Hb\left(\rho_D + \rho_m + \frac{\rho_m^2}{\rho_D + \rho_m}\right)$) into the context of the holographic Ricci dark energy model (HRDE) defined by Eqs. (15) and (16) and extended holographic Ricci dark energy model (EHRDE) defined by Eqs. (21) and (22). We used SNIa compressed Pantheon data, Baryon Acoustic

Oscillations (BAO) from BOSS DR12, Cosmic Microwave Background (CMB) of Planck 2015, fgas (gas mass fraction) and SZ/Xray (Sunyaev–Zeldovich effect and X-ray emission) as the observational data for constraining the potential free parameters of the models. For obtaining the results we employed and modified the Cosmo Hammer (EMCEE) Python package. We found that the deceleration parameter

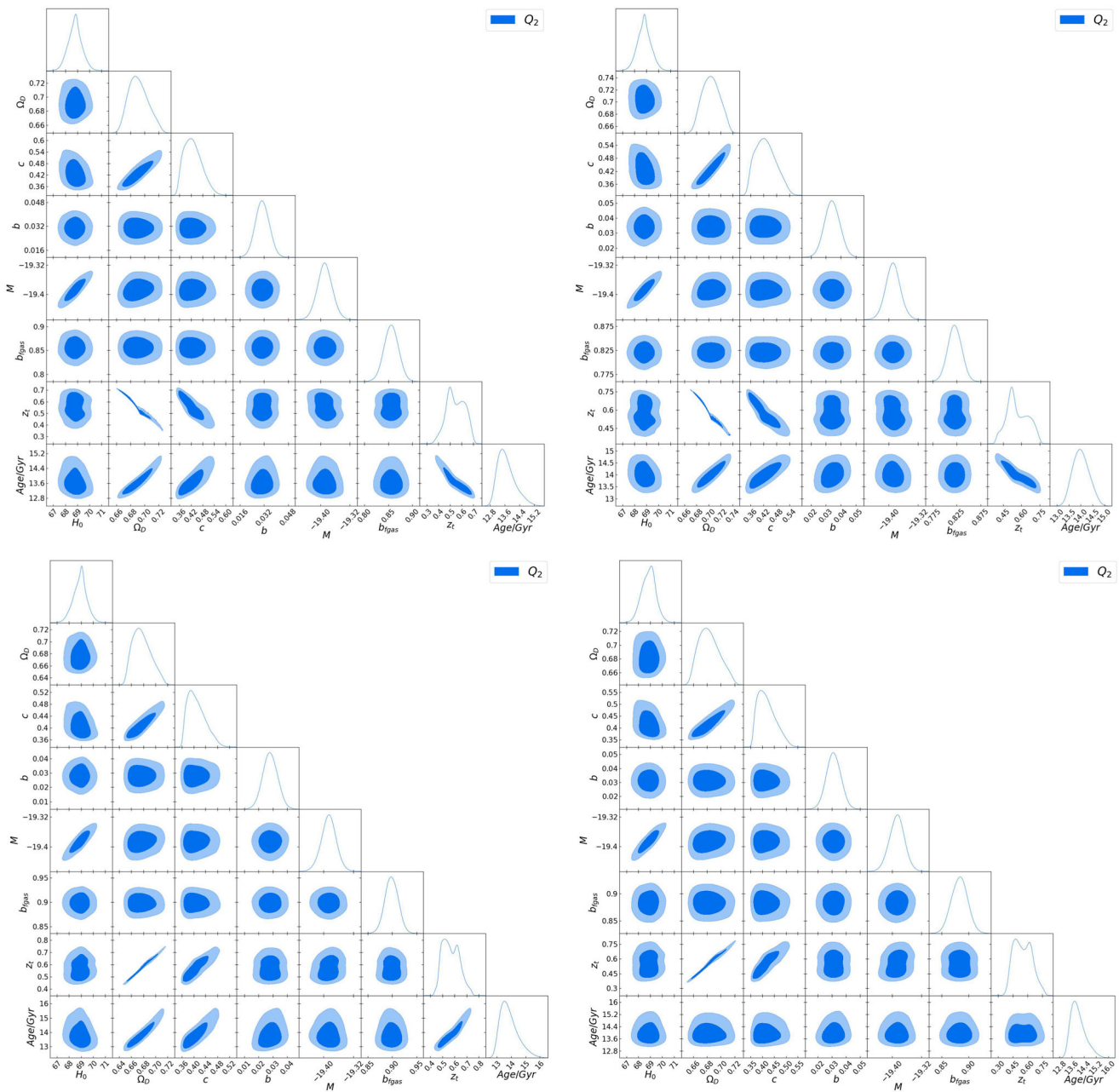

**Fig. 8** The contour maps of the HRDE (see Eqs. (15) and (16)) with four types of non-linear interaction $Q_4$, $Q_5$, $Q_6$ and $Q_7$ listed in Table 1. In this figure $H_0$ is the Hubble parameter, $\Omega_D$ is the dark energy density, $\alpha = c^2$ is the dimensionless parameter, $b$ is the coupling constant, $M$ is the nuisance parameter of SNIa data, $b_{fgas}$ is the nuisance parameter of fgas mass fraction data, $z_t$ is the transition redshift and *Age* is the age of the Universe for the HRDE model. The best-fitted values of these parameters are listed in Table 3

for all considered types of interaction, both linear and non-linear (see Table 1), shows the corresponding Universe is expanding with an accelerating rate and is in good agreement with Planck 2015 data. In addition, according to the Figs. 1 and 2 related to the deceleration parameter, it has been observed that the lower and upper bounds of $1\sigma$ confidence level for both interacting and non-interacting HRDE and EHRDE models enter the accelerating era within the redshift range $z = (0.4, 0.8)$. Using Brent's method we also obtained the transition redshift with good compatibility with recent studies, in this case, $0.4 < z_t < 1$. Studying the jerk parameter, we observed that both models cross the $\Lambda$CDM line ($j_0 = 1$) within the range of $1\sigma$ confidence level. It has been observed that the EHRDE model is closer to $j_0 = 1$ in comparison to HRDE model. By employing Bayesian Inference and obtaining the best value of parame-

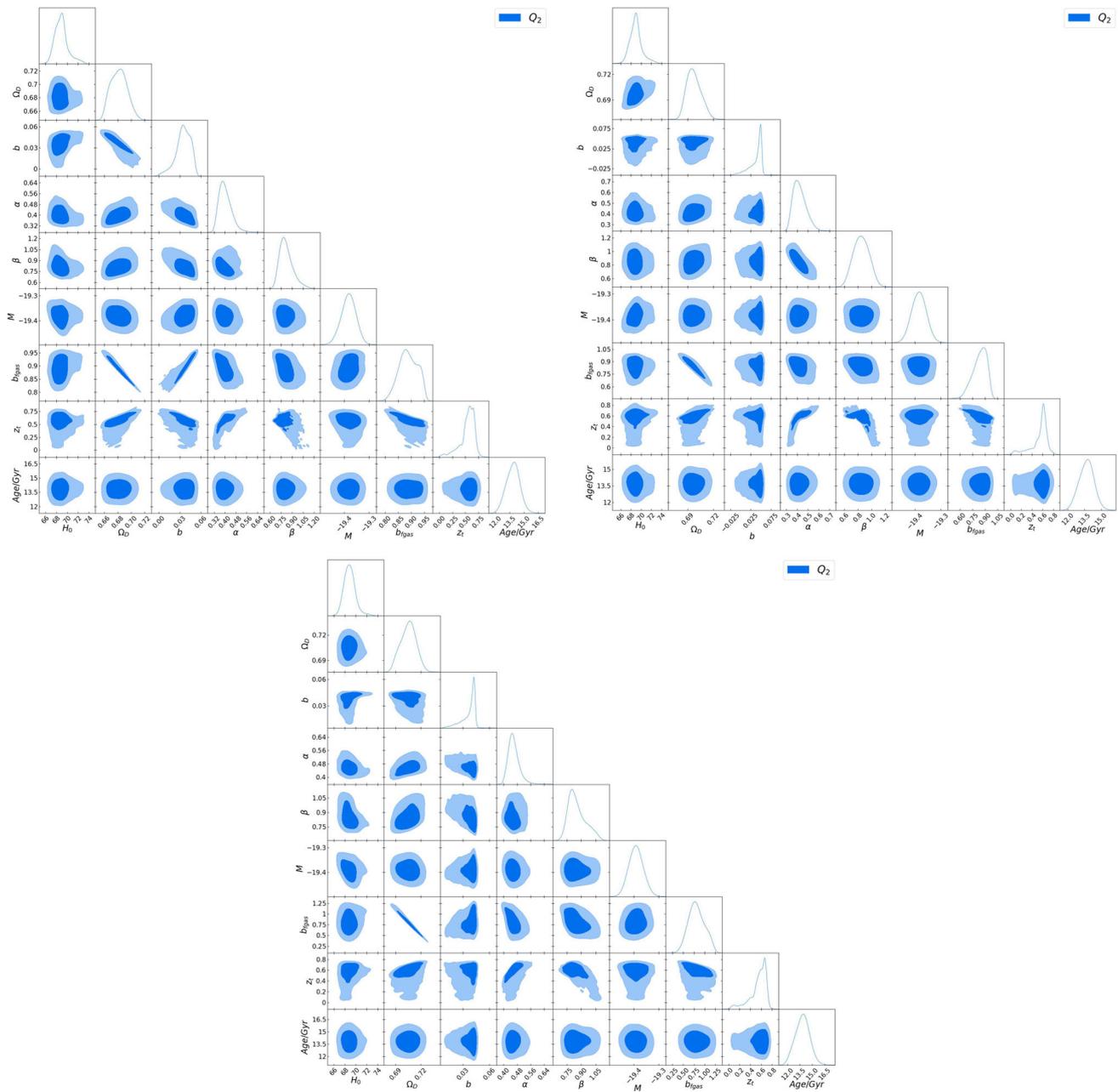

**Fig. 9** The contour maps of the EHRDE (see Eqs. (21) and (22)) with three types of linear interaction $Q_1$, $Q_2$ and $Q_3$ listed in Table 1. In this figure $H_0$ is the Hubble parameter, $\Omega_D$ is the dark energy density, $\alpha$ and $\beta$ are the dimensionless parameters, $b$ is the coupling constant, $M$ is the nuisance parameter of SNIa data, $b_{fgas}$ is the nuisance parameter of fgas mass fraction data, $z_t$ is the transition redshift and $Age$ is the age of the Universe for the EHRDE model. The best-fitted values of these parameters are listed in Table 4

ters for the $\Lambda$CDM as the reference ($H_0 = 68.5846^{+0.7970}_{-0.8015}$, $\Omega_D = 0.6968^{+0.0174}_{-0.0173}$) we found that

1. According to the value of $\Delta \ln B_{ij}$ the Bayesian Inference method shows strong evidence against the different types of HRDE whether interacting or non-interacting while the $\Lambda$CDM as the reference model is preferred over the HRDE models. Indeed with this situation, it can be also mentioned that the HDE models have been proposed because of the fundamental problems of the $\Lambda$CDM model mentioned in our discussion concerning the motivation having alternative dark energy models.

2. Within the context of Bayesian Inference, among the seven type of interactions we can pinpoint two of them

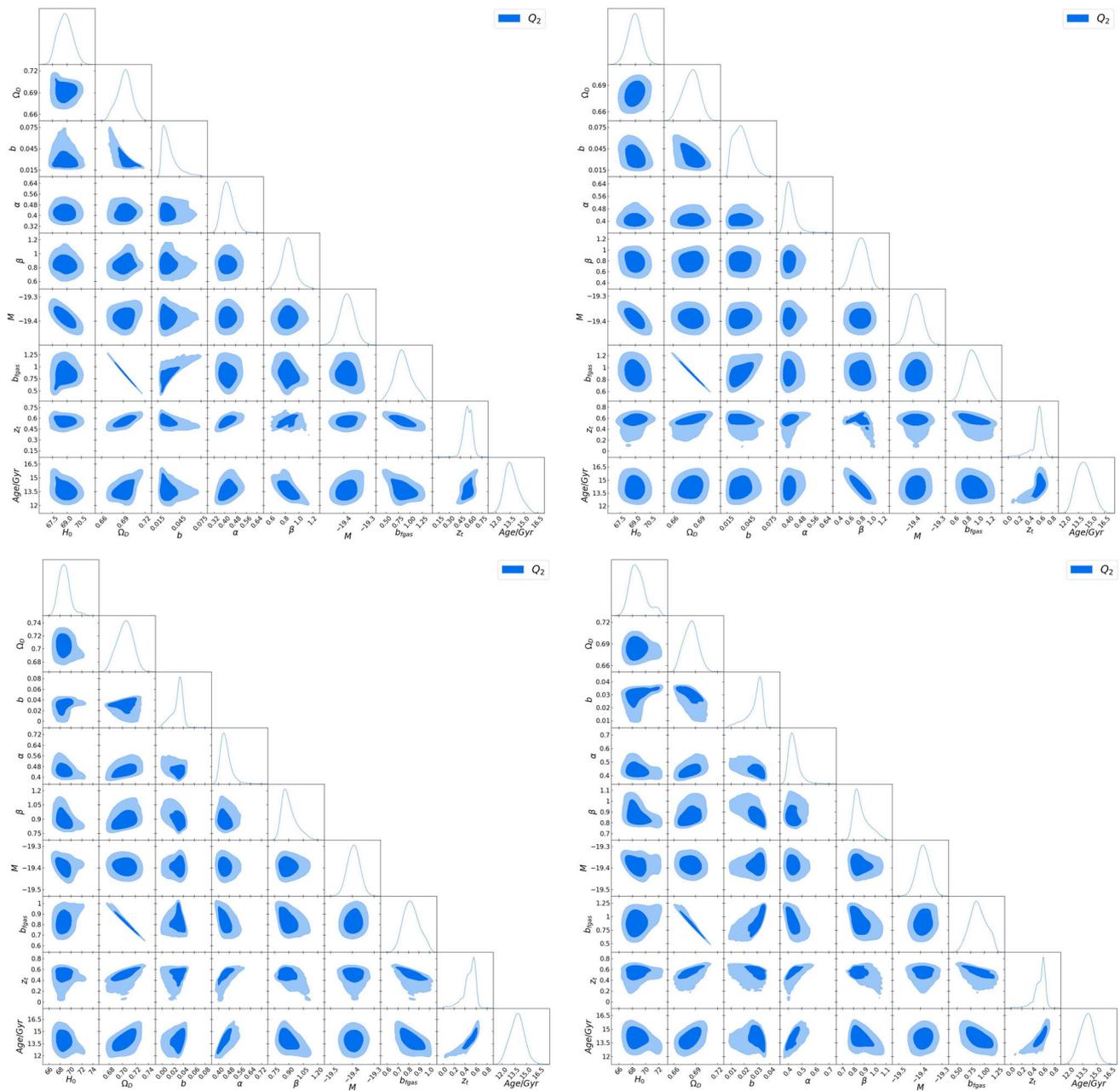

**Fig. 10** The contour maps of the EHRDE (see Eqs. (21) and (22)) with four types of non-linear interaction $Q_4$, $Q_5$, $Q_6$ and $Q_7$ listed in Table 1. In this figure $H_0$ is the Hubble parameter, $\Omega_D$ is the dark energy density, $\alpha$ and $\beta$ are the dimensionless parameters, $b$ is the coupling constant, $M$ is the nuisance parameter of SNIa data, $b_{fgas}$ is the nuisance parameter of fgas mass fraction data, $z_t$ is the transition redshift and *Age* is the age of the Universe for the EHRDE model. The best-fitted values of these parameters are listed in the Table 5

($Q_3 = 3Hb\rho_m$ and $Q_5 = 3Hb(\rho_m + \frac{\rho_m^2}{\rho_D + \rho_m})$) to be the best models.

Using a modified version of the CAMB package we observed the tendency of all models to small $\ell$ or large angle scale in power spectrum of the cosmic microwave background anisotropy and also show an as high as 20% degree of the mat-ter power spectrum's suppressing. Furthermore, we found that $Q_2$ and $Q_3$ for both HRDE and HERDE are the closest models to $\Lambda$CDM.

Finally, using the combination of observational data we fitted the free parameters of the models. We observed that the cosmological parameters of the HRDE and HERDE for all linear and non-linear interactions have good agreement with the latest obtained values of the cosmological parame-ters. Our results demonstrated that the Hubble constant value is in range of $H_0 = [0.67, 71]$ having good consistency with the recent works on observational data. The obtained value of dark energy density for all models is in good agreement with the latest Planck data. However, the model $Q_4$ with $\Omega_D = 0.6911^{+0.0121}_{-0.0161}$ for HRDE and $\Omega_D = 0.6914^{+0.0112}_{-0.0091}$ for EHRDE showed more compatibility. It worth to mention that adding two categories of galaxy clusters data namely, fgas (gas mass fraction) and SZ/X-ray (Sunyaev–Zeldovich effect and X-ray emission) did not change significantly the results compared to the previous works (mentioned in the discussion) on phenomenological interactions and also the HRDE model. The HRDE models remain unsupported by observational data and the best interaction model is the linear interaction ($Q = 3Hb\rho_m$). We also found that the depletion factor of fgas data $b_{fgas}$ should be constrained. The results of the models are very sensitive to this parameter and the assumption of $b_{fgas}$ as a fixed parameter could result in having a different value for $\Omega_D$ and even the age of the Universe. In conclusion, we can note that the new non-linear interactions (studied in this work) are reliable for further study and compatible with the latest observational data. The results showed that the galaxy clusters data namely, fgas (gas mass fraction) and SZ/X-ray (Sunyaev–Zeldovich effect and X-ray emission) can play a rational role in constraining the cosmological parameters.

It can be mentioned that for the deep understanding of phenomenological interactions, particularly the non-linear ones, more investigations should be done. Thus, for the future works, we would like to study the dynamical system methods for understanding the behavior of the non-linear interactions in the late time. We also are going to check how much these types of interactions are successful to alleviate the coincidence problem. In addition, the perturbation analysis compares to the gravitational lenses and the Large Scale Structure can be performed.

**Acknowledgements** The authors would like to thank the referee for insightful comments. Martiros Khurshudyan is supported in part by a CAS President's International Fellowship Initiative Grant (no. 2018PM0054) and the NSFC (no. 11847226). D.-f Zeng's work is supported by NSFC grant (no. 11875082).

**Data Availability Statement** This manuscript has no associated data or the data will not be deposited. [Authors' comment: The Raw Python Data used to support the findings of this study are available from the corresponding author upon request.]

## Appendix A

### A.I Compressed pantheon supernovae data

For the supernova type Ia (SNIa), we use 40 binned data points of the recent proposed Pantheon data with the range of redshift $z = [0.014, 1.62]$ [124]. We use the systematic covariance $C_{sys}$ for a vector of binned distances

$$C_{ij,sys} = \sum_{n=1}^{i} \left( \frac{\partial \mu_i}{\partial S_n} \right) \left( \frac{\partial \mu_j}{\partial S_n} \right) (\sigma_{S_k}) \tag{33}$$

in which the summation is over the $n$ systematics with $S_n$ and its magnitude of its error $\sigma_{S_n}$. According to $\triangle\mu = \mu_{data} - M - \mu_{obs}$ in which $M$ is a nuisance parameter we can write the $\chi^2$ relation for Pantheon SNIa data as

$$\chi^2_{Pantheon} = \triangle\mu^T \cdot C^{-1}_{Pantheon} \cdot \triangle\mu \tag{34}$$

Note that the $C_{Pantheon}$ is the summation of the systematic covariance and statistical matrix $D_{stat}$ having a diagonal component. The complete version of full and binned Pantheon supernova data can be found in the online source.[4]

### A.II Baryon acoustic oscillations data

We use the BOSS DR12 including six measured data points as the latest observational data for BAO [125]. The $\chi^2_{BAO}$ can be explained as

$$\chi^2_{BAO} = X^T \cdot C^{-1}_{BAO} \cdot X, \tag{35}$$

where $X$ for six data points is

$$X = \begin{pmatrix} \frac{D_M(0.38)r_{s,fid}}{r_s(z_d)} - 1512.39 \\ \frac{H(0.38)r_s(z_d)}{r_s(z_d)} - 81.208 \\ \frac{D_M(0.51)r_{s,fid}}{r_s(z_d)} - 1975.22 \\ \frac{H(0.51)r_s(z_d)}{r_s(z_d)} - 90.9 \\ \frac{D_M(0.61)r_{s,fid}}{r_s(z_d)} - 2306.68 \\ \frac{H(0.51)r_s(z_d)}{r_s(z_d)} - 98.964 \end{pmatrix}, \tag{36}$$

and $r_{s,fid} = 147.78$ Mpc is the sound horizon of fiducial model, $D_M(z) = (1+z)D_A(z)$ is the comoving angular

---

[4] https://archive.stsci.edu/prepds/ps1cosmo/index.html.

diameter distance. The covariance matrices can be found at the MontePython online files.[5]

### A.III Cosmic microwave background data

Discovering the expansion history of the Universe, we check Cosmic Microwave Background (CMB). For this, we use the data of Planck 2015 [90]. The $\chi_{CMB}^2$ function may be explained as

$$\chi_{CMB}^2 = q_i - q_i^{data} Cov_{CMB}^{-1} \left( q_i, q_j \right), \tag{37}$$

where $q_1 = R(z_*)$ is the shift parameter, $q_2 = l_A(z_*)$ in the acoustic scale, $q_3 = \omega_b$ is the density of baryonic matter and $Cov_{CMB}$ is the covariance matrix [90]. The CMB data of Planck 2015 are

$$q_1^{data} = 1.7382, \tag{38}$$
$$q_2^{data} = 301.63, \tag{39}$$
$$q_3^{data} = 0.02262. \tag{40}$$

The reader should notice that the usage of CMB data does not provide the full Planck information, but it is an optimum way of studying a wide range of dark energy models.

### A.IV Galaxy clusters' data

This method has an explicit dependency on the diameter angular distance $d_A$ of the gas mass fraction data $f_{gas}$ from the galaxy clusters. In this technique, we may consider that the baryonic fraction of the galaxy clusters proportionates to the global fraction of baryonic and dark matter. The gas mass fraction can be defined as

$$f_{gas} = \frac{M_{gas}}{M_t} \tag{41}$$

in which $M_{gas}$ is the gas mass of X-ray and $M_t$ is the total gravitational mass of the galaxy clusters. It is possible to explain the equation above according to $d_A$ [126]

$$f_{gas}^{\Lambda CDM} = \frac{b\Omega_b}{1 + 0.19\sqrt{h}\Omega_m} \left( \frac{d_A^{\Lambda CDM}}{d_A} \right)^{1.5} \tag{42}$$

in which $f_{gas}$ is observational gas mass fraction data [127], $f_{gas}^{\Lambda CDM}$ is the gas mass fraction of the cosmology models (Here HRDE Models) compared to $\Lambda CDM$ as the reference model and $b$ is the depletion component which is the key factor of relation between the baryonic fraction in the galaxy clusters and the mean cosmic value [110]. We use 42 measured data points in range of $z = [0.05, 1.1]$ [127] and we

may write the $\chi_{fgas}^2$ as

$$\chi_{fgas}^2 = \sum_{n=1}^{42} \left( \frac{f_{gas}^{\Lambda CDM} - f_{gas}^{th}}{\sigma_n} \right)^2 + \left( \frac{\Omega_b h^2 - 0.0214}{0.002} \right)^2$$
$$+ \left( \frac{h^2 - 0.072}{0.08} \right)^2 + \left( \frac{b - 0.824}{0.089} \right)^2 \tag{43}$$

For SZ/Xray data, we use 25 measured data points of angular diameter distance ($d_{A,c}$) from galaxy clusters [128]. This method can be related to observing the galaxy clusters. The processes of sudden turbulence and compaction in Intra-clusters Medium cause the temperature to rise and by the Sunyaev–Zeldovich (SZ) effect and X-ray emission the galaxy clusters can be observed [129]. Using the SZ effect and X-ray emission of galaxy clusters it can be possible to measure the diameter angular distance ($d_A$) of the clusters [130]. An error $\sigma_{de}$ is considered to each measurement which is derived by the combination of the uncertainties in the galaxy clusters and the statistical along with systematic errors. The usage of the statistical errors stems from galaxy clusters' asphericity which is among the SZ point sources and the kinetic SZ effect [131–133]. The $\chi^2$ for this procedure compared to the diameter angular distance can be written as

$$\chi_{SZ/Xray}^2 = \sum_{n=1}^{25} \left( \frac{d_A - d_{A,c}}{\sigma_{dc}} \right)^2 \tag{44}$$